\documentclass[prd,twocolumn,showpacs,showkeys,floatfix]{revtex4}
\usepackage{amsmath,amssymb}
\usepackage{latexsym}
\usepackage{bm}

\usepackage{hyperref}
\usepackage[ansinew]{inputenc}
\begin{document}
\title{COMPLEXITY AND SIMPLICITY OF SELF--GRAVITATING FLUIDS}

\author{L. Herrera}
\email{lherrera@usal.es}
\affiliation{Instituto Universitario de F\'isica
Fundamental y Matem\'aticas, Universidad de Salamanca, Salamanca 37007, Spain. \\
lherrera@usal.es}

\begin{abstract}
We review a recently proposed definition of complexity  of the structure of self--gravitating fluids \cite{ch1}, and the criterium to define the simplest mode of their evolution \cite{ch2,rg5}. We analyze the origin of these concepts and their possible applications in the study of gravitation collapse. We start by considering the static spherically symmetric case, extending next the study to static axially symmetric case \cite{cax}.
Afterward  we consider the non--static spherically symmetric case. Two possible modes of evolution are proposed to be the simplest one. One is the homologous condition  as defined in \cite{ch2}, however, as was shown later on in \cite{rg5},  it may be useful to relax this last condition to enlarge the set of possible solutions, by adopting the so-called quasi-homologous condition.
As another example of symmetry, we  consider fluids endowed with hyperbolical symmetry. Exact solutions for static fluid distributions  satisfying  the condition of minimal complexity are presented \cite{rg4}. 
An extension of the complexity factor as defined in \cite{ch1}, to the vacuum solutions of the Einstein equations represented by the Bondi metric  is discussed \cite{rg7}. A complexity hierarchy is established in this case, ranging from the Minkowski spacetime (the simplest one) to gravitationally radiating systems (the most complex).
Finally we propose  a list of questions which, we believe, deserve to be treated in the future
\end{abstract}
\date{today}

\maketitle
\section{Introduction}
\label{sec:1}

In recent decades many efforts have been made towards a rigorous definition of complexity in different branches of science (see \cite{0,1,2,1bis,2bis,2bisbis,3,3bis,3bisbis,4,4bis,6bis}  and references therein).  However, despite all the work done so far,  there is not yet a consensus on a precise definition. 

The reason behind such interest stems from the fact that at  least at  an  intuitive level,  complexity,  no matter how we define  it, is a physical concept   deeply intertwined with fundamental aspects of   the system. In other words, we expect that a suitable definition of  complexity   of the system  could allow us to infer relevant conclusions  about its behavior.

Therefore, it is of utmost relevance to provide a precise definition of an observable quantity which allows measurement of such an important property of the system. Thus, when dealing with a situation that intuitively is judged as ``complex'', we have to be able   to quantify this complexity  by defining an observable measuring it. 

Among  the many definitions that have been proposed so far, most of them resort to concepts
such as information and entropy (see for example \cite{3, 4}), and are based on the intuitive idea that  complexity should, somehow, measure a basic property  related to 
  the  structural characteristics of the   system.

This Chapter  is devoted to the review of the concept of complexity introduced in \cite{ch1} for self-gravitating systems, and its applications under variety of circumstances.

An extension of the definition of complexity based on the work developed by L\'opez-Ruiz and collaborators~\cite{3, 4}  has already been proposed  for self-gravitating systems in \cite{5,6,7,10}. 

However, such a poposal  suffers from two drawbacks, the most important of which is the fact that it  only involves the energy density, ignoring the role of stresses.  This motivated the introduction of  a quite different definition which was proposed in~\cite{ch1} for the static spherically symmetric case, and extended further in \cite{ch2} to the general full dynamic case.

The definition given in \cite{ch1}, although intuitively associated with the very concept of ``structure'' within the fluid distribution,  is not related (at least directly) to information or disequilibrium; rather it stems from the basic assumption that the simplest system (or at least one of them) is represented by the homogeneous fluid with isotropic pressure. Having assumed this conjecture for  a vanishing complexity system, the very definition of complexity  emerges in  the development of  the fundamental theory of self-gravitating compact objects, in the context of general relativity.

The variable responsible for measuring complexity, which we call the complexity factor, appears in the orthogonal splitting of the Riemann tensor, and the justification for such a proposition, roughly speaking, is as follows (details are given in the next section).

Once the point of view that for a static fluid distribution, the simplest system is represented by a homogeneous (in the energy density), locally  isotropic fluid (principal stresses equal),   then it is reasonable to assign  zero value of  the complexity factor to such a distribution. Next, let us  recall the concept of  Tolman mass~\cite{11}, which may be interpreted as the ``active''  gravitational mass, and may be expressed, for an arbitrary distribution,  through its value for the zero-complexity case plus two terms depending on the energy density inhomogeneity and pressure anisotropy, respectively.  These latter terms in turn may be expressed through a single scalar function that we call the complexity factor. It obviously vanishes when the fluid is homogeneous in the energy density,  and isotropic in pressure, but also may vanish when the two terms  containing density inhomogeneity and anisotropic pressure cancel each other out. Thus, as in~\cite{3}, vanishing complexity  may correspond to very different systems.

When dealing with time-dependent  systems, we face two different problems; on the one hand, we have to generalize the concept of complexity  of the structure of the fluid distribution to time-dependent dissipative fluids, and on the other hand we also have to evaluate the complexity of the patterns of evolution and propose what we consider to be the simplest of them. 

In~\cite{ch2} it was shown that is reasonable to assume that the complexity factor for the structure of the fluid distribution is the same scalar function  as for the static case, which now includes the dissipative variables. As for the simplest pattern of evolution, it was shown that the homologous condition  characterizes the simplest possible mode. However, as was shown later on,  it may be useful to relax this last condition to enlarge the set of possible solutions, by adopting the so-called quasi-homologous condition \cite{rg5}.

 The axially symmetric static case has been considered  in~\cite{cax}, while  some particular cases of cylindrically symmetric fluid distributions have been studied in~\cite{ch, cil}. Further applications of the concept of complexity as defined in~\cite{ch1} may be found in~\cite{rg,rg1,rg2,rg3,rg4}. Always within the context of general relativity, exact solutions for static fluid distributions endowed with hyperbolical symmetry and satisfying  the condition of minimal complexity were presented in~\cite{rg4,c13,c22}, while  dynamic solutions endowed with hyperbolical symmetry and satisfying  the condition of minimal complexity were obtained in~\cite{rg6}, such solutions evolve in the so-called quasi-homologous regime. 

 The concept of complexity as defined in~\cite{ch1} has also been extended   to other theories of gravity in~\cite{ot1,ot2,ot3,ot4,n1,ot5,ot6,ot7,ot8,ot9,ot10,ot11,ot12,ot13,ot14,ot15,ot16,ot17,ot18,ot19,ot20}. More recent developments  of this subject within a variety of contexts may be found in \cite{c20,c1,c6,c9,c11,c15,c10,c16,c5,c17,c18,c8,c12,c23,c23bis,c23bc,Al,c2,c24,c14,c19,c21,c21b,c4,c3,c7,c7b,c7c,c7d,rin1,rin2,ja1} and references therein.

All the above-mentioned cases concern fluid distributions (which eventually may be charged); however, the vacuum case has barely been treated. The only known example so far is the extension of the complexity factor as defined in~\cite{ch1} to the vacuum solutions of the Einstein equations represented by the Bondi metric~\cite{rg7}. A complexity hierarchy may  be established in this case, ranging from the Minkowski spacetime (the simplest one) to gravitationally radiating systems (the most complex).

\section{Defining complexity for static self--gravitating systems}
\label{sec:2}
Let us start by introducing the notion of complexity of a self--gravitating fluid as represented by a variable (more than one for non--spherical configurations) aimed to  measure the degree of ``complexity'' of the fluid distribution. At this level of generality, due to the vagueness of the concept of  ``complexity'',  the problem appears to be quite difficult to solve, accordingly let's consider instead a much simpler question, namely: what is the simplest fluid configuration we can envisage? 

In order to answer to the above question, let us first restrict the discussion to static and spherically symmetric distributions. Under these conditions, the fluid is described by  the energy density and the pressure. The simplest energy density distribution corresponds to  an incompressible fluid (a constant energy density) whereas the simplest pressure distribution corresponds to an isotropic fluid (principal stresses equal). Indeed, it should be noticed that since for a bounded configuration the pressure vanishes at the boundary surface, it cannot be homogeneous. 

Thus, based on these very simple considerations we shall assume that the simplest structure (at least one of them) of a static spherically symmetric fluid corresponds to a homogeneous (in the energy density) and isotropic (in the pressure) fluid. We shall now see how this assumption leads us to a reliable definition of complexity.

\subsection*{The static  spherically symmetric case}
We consider spherically symmetric distributions of  static 
fluid, which for the sake of completeness we assume to be locally anisotropic and  bounded by a
spherical surface $\Sigma$.

\noindent
The line element is given in Schwarzschild--like  coordinates by

\begin{equation}
ds^2=e^{\nu} dt^2 - e^{\lambda} dr^2 -
r^2 \left( d\theta^2 + \sin^2\theta d\phi^2 \right),
\label{metrica}
\end{equation}

\noindent
where $\nu(r)$ and $\lambda(r)$ are functions of their arguments. We
number the coordinates: $x^0=t; \, x^1=r; \, x^2=\theta; \, x^3=\phi$.

The energy--momentum tensor  corresponding to our  fluid distribution may be written as 
\begin{equation}
T^{\mu}_{\nu}= \mu u^{\mu}u_{\nu}-  P
h^{\mu}_{\nu}+\Pi ^{\mu}_{\nu},
\label{24'}
\end{equation}
with 
\begin{eqnarray}
\Pi^{\mu}_{\nu}&=&\Pi(s^{\mu}s_{\nu}+\frac{1}{3}h^{\mu}_{\nu});\quad
P=\frac{P_{r}+2P_{\bot}}{3}\nonumber \\ &&\Pi=P_{r}-P_{\bot};\quad h^\mu_\nu=\delta^\mu_\nu-u^\mu u_\nu,
\label{varios}
\end{eqnarray}
where $\mu, P_r, P_\bot, \Pi_{\mu \nu}, u^\mu$ denote the energy density, the radial pressure, the tangential pressure, the anisotropic tensor and the four--velocity respectively.

The vector   $s^\mu$  is  defined by
\begin{equation}
s^{\mu}=(0,e^{-\frac{\lambda}{2}},0,0)\label{ese},
\end{equation}

whereas  the four--velocity  vector is  given by:
\begin{equation} 
u^{\mu}=(e^{-\frac{\nu}{2}},0,0,0)\label{u},
\end{equation}

with the  properties
$s^{\mu}u_{\mu}=0$,
$s^{\mu}s_{\mu}=-1$.
 
 From (\ref{u})  we can calculate the four acceleration, $a^\alpha=u^\alpha_{;\beta}u^\beta$, whose only non--vanishing component is:
\begin{equation}
 a_1=-\frac{\nu
^{\prime}}{2},
\label{15a}
\end{equation}
where prime denotes derivative with respect to $r$.

\noindent
At the exterior of the fluid distribution the spacetime is described by the  Schwarzschild metric

\begin{equation}
ds^2= \left(1-\frac{2M}{r}\right) dt^2 - \frac{dr^2}{ \left(1-\frac{2M}{r}\right)} -
r^2 \left(d\theta^2 + \sin^2\theta d\phi^2 \right).
\label{Vaidya}
\end{equation}

\noindent
In order to match smoothly the two metrics above on the boundary surface
$r=r_\Sigma=constant$, we  require the continuity of the first and the second fundamental
forms across that surface, producing 
\begin{equation}
e^{\nu_\Sigma}=1-\frac{2M}{r_\Sigma},
\label{enusigma}
\end{equation}
\begin{equation}
e^{-\lambda_\Sigma}=1-\frac{2M}{r_\Sigma},
\label{elambdasigma}
\end{equation}
\begin{equation}
\left[P_r\right]_\Sigma=0,
\label{PQ}
\end{equation}
where, from now on, subscript $\Sigma$ indicates that the quantity is
evaluated on the boundary surface $\Sigma$.

Eqs. (\ref{enusigma}), (\ref{elambdasigma}), and (\ref{PQ}) are the necessary and
sufficient conditions for a smooth matching of the two metrics (\ref{metrica})
and (\ref{Vaidya}) on $\Sigma$.

\noindent
Next, let us recall that the Tolman mass for a spherically symmetric static distribution 
of matter is given by  \cite{11}

\begin{eqnarray}
m_T =    4\pi \int^{r_\Sigma}_{0}{r^2 e^{(\nu+\lambda)/2} 
\left(T^0_0 - T^1_1 - 2 T^2_2\right) dr},
\label{Tol}
\end{eqnarray}

\noindent
 which we shall extend for any sphere  of 
radius $r$, completely inside $\Sigma$, as 

\begin{eqnarray}
m_T =   4\pi \int^{r}_{0}{\tilde r^2 e^{(\nu+\lambda)/2} 
\left(T^0_0 - T^1_1 - 2 T^2_2\right) d\tilde r}\nonumber .
\label{Tolin}
\end{eqnarray}

\noindent
This extension of the global concept of energy to a local level 
 is suggested by the conspicuous role played by 
$m_T$ as the ``active gravitational mass''.

 Indeed, as it follows from (\ref{15a}), the gravitational acceleration ($a=-s^\nu a_\nu$) of a test particle, 
instantaneously at rest in a static gravitational field,  is given by 

\begin{equation}
a = \frac{e^{- \lambda/2} \, \nu'}{2} =  \frac{e^{-\nu/2}m_T}{r^2} .
\label{a}
\end{equation}

Another expression for $m_T$, 
which appears to be more suitable for the discussion  is (see \cite{13, 14} for details, but notice  slight changes in notation)

\begin{widetext}
\begin{eqnarray}
m_T  =  (m_T)_\Sigma \left(\frac{r}{r_\Sigma}\right)^3  -  r^3 \int^{r_\Sigma}_r{e^{(\nu+\lambda)/2} \left[\frac{8\pi}{\bar r} 
\left(P_\bot - P_r \right)
+ \frac{1}{\bar r^4} \int^{\bar r}_0 {4\pi \tilde{r}^3 \mu' d\tilde{r}} 
 \right] d\bar r} ,
\label{emtebisbis}
\end{eqnarray}
\end{widetext}
The important point to keep in mind here is that the second  integral in  (\ref{emtebisbis}) describes the contribution of density inhomogeneity and local anisotropy of pressure to the Tolman mass. 

Next, using the  orthogonal splitting of the Riemann tensor first considered by Bel \cite{bel1} (see also \cite{GP}), let us introduce the tensor
\begin{equation}
Y_{\alpha \beta}=R_{\alpha \gamma \beta \delta}u^{\gamma}u^{\delta},
\label{electric}
\end{equation}
which can be splitted in terms of its trace and  the corresponding trace--free tensor, as (see \cite{sc} for details) 

\begin{equation}
TrY\equiv Y_T=4\pi(\rho+3P_r-2\Pi),
\label{esnV}
\end{equation}
and
\begin{equation}
Y_{<\alpha \beta>}=Y_{TF}(s_\alpha s_\beta+\frac{h_{\alpha \beta}}{3}),
\label{esnVI}
\end{equation}
with  
\begin{equation}
Y_{TF}\equiv (4\pi  \Pi+E)=8\pi \Pi- \frac{4\pi}{r^3} \int^r_0{\tilde r^3 \mu' d\tilde r},
\label{defYTFa}
\end{equation}

where $E$ is the scalar defining the electric part of the Weyl tensor $E_{\alpha \beta}=C_{\alpha \gamma \beta
\delta}u^{\gamma}u^{\delta}$,
(the magnetic part vanishes identically),  which may  be written as 
\begin{equation}
E_{\alpha \beta}=E (s_\alpha s_\beta+\frac{1}{3}h_{\alpha \beta}),
\label{52bisx}
\end{equation}
with
\begin{equation}
E=-\frac{e^{-\lambda}}{4}\left[ \nu ^{\prime \prime} + \frac{{\nu
^{\prime}}^2-\lambda ^{\prime} \nu ^{\prime}}{2} -  \frac{\nu
^{\prime}-\lambda ^{\prime}}{r}+\frac{2(1-e^{\lambda})}{r^2}\right],
\label{defE}
\end{equation}
satisfying the following properties:
 \begin{eqnarray}
 E^\alpha_{\,\,\alpha}=0,\quad E_{\alpha\gamma}=
 E_{(\alpha\gamma)},\quad E_{\alpha\gamma}u^\gamma=0.
  \label{propE}
 \end{eqnarray} 

Using (\ref{defYTFa})  in (\ref{emtebisbis}) we get
 
\begin{eqnarray}
m_T  =  (m_T)_\Sigma \left(\frac{r}{r_\Sigma}\right)^3
 +  r^3 \int^{r_\Sigma}_r{\frac{e^{(\nu+\lambda)/2}}{\tilde r} Y_{TF}d\tilde r} .
\label{emtebis}
\end{eqnarray}

Thus we see that this single scalar function, $Y_{TF}$, encompasses all the modifications produced  by the energy density inhomogeneity and the anisotropy of the pressure, on the active  gravitational (Tolman) mass. More specifically, it describes how these two factors modify the value of the Tolman mass, with respect to its value for the homogeneous isotropic fluid. 

Then, following our starting conjecture stating that at least one of  the simplest configurations corresponds  to an incompressible fluid (a constant energy density) with isotropic pressure, and noticing  
that 
for such a system,  the scalar $Y_{TF}$ vanishes, it appears well justified to identify the complexity factor with $Y_{TF}$. 

The following remarks are in order at this point:
\begin{itemize}
\item The scalar $Y_{TF}$ belongs to a family of functions known as structure scalars, defined in \cite{sc} from a detailed analysis  of the orthogonal splitting of the Riemann tensor.
\item As is apparent from (\ref{a}) the Toman mass is a measure of the strength of the gravitational interaction.
\item The complexity factor  defined as above, not only vanishes for the homogeneous, isotropic fluid, where the two terms in (\ref{defYTFa}) vanish identically, but also for all configurations where the two terms in (\ref{defYTFa}) cancel each other, implying that there are a wealth of configurations satisfying the vanishing  complexity condition.
\item It is worth noticing that whereas the contribution of the pressure anisotropy to $Y_{TF}$ is local, the contribution of the density energy inhomogeneity is not.
\item For a charged  fluid  $Y_{TF}$ includes contributions from the electric charge as well (see eq.(25) in \cite{21}).
\end{itemize}

In the next section, we shall present two  examples of inhomogeneous and anisotropic fluid configurations, satisfying the vanishing complexity factor condition.

\subsection{Fluid distributions with vanishing complexity factor}
The vanishing complexity factor condition, according to (\ref{defYTFa}) reads

\begin{equation}
 \Pi=\frac{1}{2r^3} \int^r_0{\tilde r^3 \mu' d\tilde r}.
\label{vcfc}
\end{equation}

However, since  the Einstein equations for  a spherically symmetric static, anisotropic fluid  form  a system of three ordinary differential equations for five unknown functions ($\nu, \lambda, \mu, P_r, P_\bot$),  the condition $Y_{TF}=0$ is not enough to close the system and we  need still one condition in order to solve it. 
Just for the sake of illustration, we shall propose two  examples in the next subsections.

\subsubsection{The Gokhroo and Mehra ansatz}
A family of anisotropic spheres has been found in \cite{22}, which leads to physically satisfactory models for compact objects.

These models are obtained from an  ansatz on the form of the metric function $\lambda$ which reads

\begin{equation}
e^{-\lambda}=1-\alpha r^2+\frac{3K\alpha r^4}{5r_\Sigma^2},
\label{lgm}
\end{equation}
producing, 
\begin{equation}
\mu=\mu_0\left(1-\frac{Kr^2}{r_\Sigma^2}\right),
\label{mugm}
\end{equation}
and
\begin{equation}
m(r)=\frac{4\pi\mu_0r^3}{3}\left(1-\frac{3Kr^2}{5r_\Sigma^2}\right),
\label{mgm}
\end{equation}

where $K$ is a constant in the range $(0, 1)$, $\alpha=\frac{8\pi \mu_0}{3}$, and $m(r)$ is the mass function, defined by 
 \begin{equation}
1-e^{-\lambda}=\frac{2m}{r},
\label{rieman}
\end{equation}
or, equivalently 
\begin{equation}
m = 4\pi \int^{r}_{0} \tilde r^2 \mu d\tilde r.
\label{m}
\end{equation}

Then using the field equations it can be shown that the line element becomes (see \cite{ch1} for details)
\begin{widetext}
\begin{equation}
ds^2=-e^{\int (2z(r)-2/r)dr}dt^2+\frac{z^2(r) e^{\int(\frac{4}{r^2
z(r)}+2z(r))dr}} {r^6(-2\int\frac{z(r)(1+\frac{\Pi (r)r^2}{8\pi})
e^{\int(\frac{4}{r^2
z(r)}+2z(r))dr}}{r^8}dr+C)}dr^2+r^2d\theta^2+r^2sin^2\theta
d\phi^2. \label{metric2}
\end{equation}
\end{widetext}
where $C$ is a constant of integration, and 

\begin{equation}
e^{\nu (r)}=e^{\int (2z(r)-2/r)dr}.
\label{v1}
\end{equation}

For  the physical variables  we have

\begin{equation}
4\pi P_r=\frac{z(r-2m)+m/r-1}{r^2}\label{Pr},
\end{equation}
\noindent 

 \begin{equation}
4\pi \mu =\frac{m^{\prime}}{r^2}\label{rho},
\end{equation}
and 
\begin{equation}
4\pi P_\bot=(1-\frac{2m}{r})(z^{\prime}+z^2-\frac{z}{r}+\frac{1}{r^2})+z(\frac{m}{r^2}-\frac{m^{\prime}}{r}).
\label{Pbot}
\end{equation}

The so obtained solution is regular at the origin, and satisfies the conditions $\mu>0$, $\mu>P_r, P_\bot$.

Also, to avoid singular behaviour of physical variables on the boundary of the source ($\Sigma$), the solution should satisfy the Darmois conditions on the boundary  (\ref{enusigma}), (\ref{elambdasigma}), (\ref{PQ}). 

\subsubsection{The polytrope with vanishing complexity factor}
The polytropic equation of state plays an important role in the study of self--gravitating systems, both, in Newtonian and general relativistic astrophysics. In the isotropic pressure case such an equation of state is sufficient to integrate the field equations. However in case of an anisotropic fluid additional information is required. The study of polytropes for anisotropic matter has been considered in detail in \cite{23, 24, 25}. 

Once the polytropic equation of state is assumed, in the case of anisotropic matter, we still need an additional condition in order to solve the corresponding system of equations.
We propose to assume the vanishing   complexity factor condition as  such a complementary information.

Thus our model is characterized by
\begin{equation}P_r=K\mu^{\gamma}=K\mu^{1+1/n};\qquad Y_{TF}=0,
\label{p2}\end{equation} 
where constants $K$, $\gamma$, and $n$ are usually called   polytropic constant, polytropic exponent, and polytropic index, respectively.

It is worth mentioning that the generalization of the Newtonian polytrope to the  general relativistic case admits two possibilities. One is the equation (\ref{p2}), the other is 
\begin{equation}
P_r=K\mu_{b}^{\gamma}=K\mu_b^{1+1/n}
\label{p3}
\end{equation}
where $\mu_{b}$ denotes the baryonic (rest) mass density. The treatment of this latter  case has been described in detail in \cite{24}.

From the polytropic equation of state (\ref{p2}) we obtain two equations which read:
\begin{widetext}
\begin{equation}
\xi^2 \frac{d\Psi}{d\xi}\left[\frac{1-2(n+1)\alpha v/\xi}{1+\alpha \Psi}\right]+v+\alpha\xi^3 \Psi^{n+1}+\frac{2\Pi \Psi^{-n}\xi}{P_{rc}(n+1)} \left[\frac{1-2\alpha(n+1)v/\xi}{1+\alpha \Psi}\right]=0,\label{TOV2anis_WB}
\end{equation}
\end{widetext}
and 
\begin{equation}\frac{dv}{d\xi}=\xi^2 \Psi^n,\label{veprima2}\end{equation}
where
\begin{equation}
\alpha=P_{rc}/\mu_{c},\quad r=\xi/A,  \quad A^2=4 \pi \mu_{c}/\alpha (n+1)\label{alfa},\end{equation}

\begin{equation}\Psi^n=\mu/\mu_{c},\quad v(\xi)=m(r) A^3/(4 \pi\mu_{c}),\label{psi}\end{equation}
where subscript $c$ indicates that the quantity is evaluated at the center. At the boundary surface $r=r_\Sigma$ ($\xi=\xi_\Sigma$) we have $\Psi(\xi_\Sigma)=0$ (see \cite{25} for details).

Equations (\ref{TOV2anis_WB}), (\ref{veprima2}), form a system of two first order  ordinary differential equations for the three unknown functions: $\Psi, v, \Pi$, depending on a duplet of parameters $n, \alpha$. In order to proceed further with the modeling of a compact object,  we shall  assume the vanishing complexity factor condition, which with the notation above, reads
\begin{equation}
\frac{6 \Pi}{n\mu_c}+\frac{2\xi}{n\mu_c}\frac{d\Pi}{d\xi}=\Psi^{n-1}\xi \frac{d\Psi}{d\xi}.
\label{pol3}
\end{equation}

Now we have a system of three ordinary differential equations (\ref{TOV2anis_WB}), (\ref{veprima2}), (\ref{pol3}) for the three unknown functions $\Psi, v, \Pi$, which may be integrated for any arbitrary duplet of values of the parameters $n, \alpha$, only constrained by the physical conditions (see \cite{24} for details)
\begin{equation}
\mu>0, \qquad \alpha \Psi \leq 1, \qquad \frac{3v}{\xi^3 \Psi^n}+\alpha \Psi-1 \leq1.
\label{conditionsIII}
\end{equation}

The equations equivalent to (\ref{TOV2anis_WB}) and (\ref{pol3}), for the equation of state (\ref{p3}) read
\begin{widetext}
\begin{eqnarray}
\xi^2 \frac{d\Psi_b}{d\xi}\left[\frac{1-2(n+1)\alpha v/\xi}{1+\alpha \Psi_b}\right]+v+\alpha\xi^3 \Psi_b^{n+1}+\frac{2\Pi \Psi_b^{-n}\xi}{P_{rc}(n+1)} \left[\frac{1-2\alpha(n+1)v/\xi}{1+\alpha \Psi_b}\right]=0,\label{TOV1anis_WB}
\end{eqnarray}
\end{widetext}

\begin{equation}
\frac{6 \Pi}{n\mu_{bc}}+\frac{2\xi}{n\mu_{bc}}\frac{d\Pi}{d\xi}=\Psi_b^{n-1}\xi \frac{d\Psi_b}{d\xi}\left[1+K(n+1)\mu_{bc}^{1/n} \Psi_b\right],
\label{pol4}
\end{equation}
with $\Psi_b^n=\mu_b/\mu_{bc}$.

Further research on the application of the concept of complexity in the study of polytropes  may be found in \cite{rg2b,rg2,rg3b,rg4b,c23,c12}.

So far we have been concerned with spherically symmetric systems, we shall next consider the axially symmetric (static) case.

\subsection*{The static axially--symmetric case}

Let us now extend the concept of complexity as defined for the spherically symmetric case  to the most general axially symmetric static fluid distributions.

The reason to undertake such a task is  based on the well known fact   that  (close to the horizon) there is a  bifurcation between any finite perturbation of Schwarzschild spacetime and any Weyl solution, even when the latter is characterized by parameters arbitrarily close to those corresponding to spherical symmetry (see \cite{i1,i5,in,i3,i2,i4} and references therein for a discussion on this point). This fact in turn is related to the well known result, usually referred to as Israel theorem   \cite{israel}, stating that the only regular static and asymptotically flat vacuum spacetime possessing a regular horizon is the Schwarzchild solution, while all the others Weyl exterior solutions  exhibit singularities in the  curvature invariants (as the boundary of the source approaches the horizon).

Thus, even though  observational evidence  suggest  that deviations from spherical symmetry in compact
self-gravitating objects (white dwarfs, neutron stars), are likely to be incidental rather than basic features of these systems, it is clear that for very compact objects, deviations from spherical symmetry (no matter how small) should be studied resorting to exact solution of the Einstein equations, instead as perturbations of spherically symmetric systems.

As shown in the previous sections, the scalar intended to measure the degree of complexity for a spherically symmetric fluid distribution (the complexity factor), is  identified as one of the scalar functions (structure scalars)  which appears in the orthogonal splitting of the Riemann tensor.  More specifically, it is related to one of the scalar functions appearing in the splitting of the electric part of the Riemann tensor.

In spite of the fact that in the axially symmetric case the situation is much more complicated, and the number of structure scalars is  larger than in the spherically symmetric case, we shall proceed in a similar way  to fix the general criterium  to define the variable(s) measuring the complexity of the fluid distribution. 

Thus,  we  start by asking ourselves the same question as in the previous case, namely: which is the simplest fluid configuration?  As in the spherically symmetric case we shall assume that such a configuration corresponds to the incompressible (constant energy density), isotropic (in the pressure) spheroid. From this simple assumption, we shall see that as the obvious candidates to measure the degree of complexity of the fluid distribution, appear three of the eight structure scalars corresponding to the axially symmetric static fluid distribution. Explicit forms of these structure scalars as well as some useful differential  equations relating the inhomogeneities of the energy density to some of the structure scalars were already found in \cite{as}.

As in the spherically symmetric case, the vanishing of the three complexity factors corresponds not only to the incompressible, isotropic  spheroid, but also to a large family of solutions where the density inhomogeneity terms cancel the pressure anisotropic terms in the equations relating these variables to the complexity factors. Some of these solutions will be exhibited.

Thus, let us  consider  static and axially symmetric sources. For such a system the line element may be written in ``Weyl spherical coordinates''  (please notice that in this section we are using signature $+2$ instead of $-2$ as in the previous case, which leads to some changes in the sign of some variables), as
\begin{equation}
ds^2=-A^2 dt^2 + B^2 \left(dr^2 +r^2d\theta^2\right)+D^2d\phi^2,
\label{1b}
\end{equation}
where the coordinates $t$ and $\phi$ are adapted to the two Killing vectors admitted by our line element, and therefore the metric functions depend only on $r$ and $\theta$.

We recall that, unlike the vacuum case, the assumption of the Weyl gauge ( $R^3_3+R^0_0=0$, where $R^\alpha_\beta$ denotes the Ricci tensor), reducing   the number of independent metric functions to two, cannot be used  without loss of generality  in the interior, which explains why our line element is described in terms of three independent functions.

 In a purely locally Minkowski frame  (hereafter referred to as l.M.f.) where the first derivatives of the metric vanish (locally) \cite{Bo}, the most general  energy--momentum tensor is given by:

\begin{equation}
\widehat{T}_{\alpha\beta}= \left(\begin{array}{cccc}\mu    &  0  &   0     &   0    \\0 &  P_{xx}    &   P_{xy}     &   0    \\0       &   P_{yx} & P_{yy}  &   0    \\0       &   0       &   0     &   P_{zz}\end{array} \right) \label{3},
\end{equation}
\\
where $\mu, P_{xy}, P_{xx}, P_{yy}, P_{zz}$ denote the energy density and different stresses, respectively, as measured by our locally defined Minkowskian observer.

Also observe that  $P_{xy}= P_{yx} $ and, in general  $ P_{xx}  \neq  P_{yy}  \neq P_{zz}$.

Then transforming back to our coordinates, we obtain the components of the energy momentum tensor in terms of the physical variables as defined in the l.M.f.
\begin{eqnarray}
{T}_{\alpha\beta}&=& (\mu+P_{zz}) V_\alpha V_\beta+P_{zz} g _{\alpha \beta} +(P_{xx}-P_{zz}) K_\alpha  K_\beta\nonumber \\ &+& (P_{yy}-P_{zz}) L_\alpha L_\beta +2P_{xy} K_{(\alpha}  L_{\beta)},
\label{6}
\end{eqnarray}
with
\begin{eqnarray}
 V_\alpha=(-A,0,0,0);\quad  K_\alpha=(0,B,0,0);\nonumber \\
   L_\alpha=(0,0,Br,0); \quad S_{\alpha}=(0, 0, 0, D),
\label{7}
\end{eqnarray}
where we are considering observers at rest with respect to the fluid distribution.

Alternatively we may write the energy momentum tensor in the ``canonical'' form
\begin{eqnarray}
{T}_{\alpha\beta}&=& (\mu+P) V_\alpha V_\beta+P g _{\alpha \beta} +\Pi_{\alpha \beta},
\label{6bisax}
\end{eqnarray}
with
\begin{eqnarray}
\Pi_{\alpha \beta}&=&(P_{xx}-P_{zz})\left(K_\alpha K_\beta-\frac{h_{\alpha \beta}}{3}\right)\nonumber \\&+&(P_{yy}-P_{zz})\left(L_\alpha L_\beta-\frac{h_{\alpha \beta}}{3}\right)+2P_{xy}K_{(\alpha}L_{\beta)}
\label{6bb},
\end{eqnarray}
\begin{equation}
P=\frac{P_{xx}+P_{yy}+P_{zz}}{3}, \quad h_{\mu \nu}=g_{\mu \nu}+V_\nu V_\mu.
\label{7Pb}
\end{equation}

The anisotropic tensor may also be written as
\begin{widetext}
\begin{eqnarray}
 \Pi_{\alpha \beta}=\frac{1}{3}(2\Pi_I+\Pi_{II})\left( K_\alpha  K_\beta -\frac{ h_{\alpha
\beta}}{3}\right)+\frac{1}{3}(2\Pi _{II}+ \Pi_I)\left( L_\alpha   L_\beta -\frac{ h_{\alpha
\beta}}{3}\right)+ \Pi _{KL}\left( K_\alpha
 L_\beta+ K_\beta
 L_\alpha\right) \label{6bba},
\end{eqnarray}
\end{widetext}
with
\begin{eqnarray}
 \Pi _{KL}= K^\alpha  L^\beta T_{\alpha \beta},
 \label{7P}
\end{eqnarray}

\begin{equation}
 \Pi_I=\left(2  K^\alpha   K^\beta -   L^\alpha   L^\beta-  S^\alpha  S^\beta \right)  T_{\alpha \beta},
\label{2n}
\end{equation}
\begin{equation}
 \Pi_{II}=\left(2  L^\alpha  L^\beta - K^\alpha   K^\beta- S^\alpha   S^\beta \right) T_{\alpha \beta}.
\label{2nbis}
\end{equation}

The relationships between the above scalars and the variables $P_{xy}, P_{xx}, P_{yy}, P_{zz}$ are (besides (\ref{7Pb})), 
\begin{equation}
\Pi_2\equiv \frac{1}{3}(2\Pi_I+\Pi_{II})=P_{xx}- P_{zz},
\label{3nbis}
\end{equation}
\begin{equation}
\Pi_3\equiv \frac{1}{3}(2\Pi_{II}+\Pi_{I})=P_{yy}- P_{zz},
\label{4nbis}
\end{equation}
\begin{equation}
\Pi_{KL}=P_{xy}.
\label{5nbis}
\end{equation}
or, inversely:
\begin{equation}
 P_{zz}=P-\frac{1}{3}(\Pi_2+\Pi_3),
 \label{6nbis}
\end{equation}
\begin{equation}
 P_{xx}=P+\frac{1}{3}(2\Pi_2-\Pi_3),
\label{7nbis}
\end{equation}
\begin{equation}
P_{yy}=P+\frac{1}{3}(2\Pi_3-\Pi_2).
\label{8nbis}
\end{equation}

The explicit form of the Einstein equations as well as the conservation equations, for the line element (\ref{1b}) and the energy--momentum tensor (\ref{6bisax}), are given in the Appendix \ref{A}
\subsection{The structure scalars}
The structure scalars for our problem were calculated in \cite{as}.  For their definition we need  first to obtain the electric part of the Weyl tensor (the magnetic part vanishes identically), whose components  can be obtained directly from its definition,
\begin{equation}
E_{\mu\nu}=C_{\mu\alpha\nu\beta}\,V^\alpha\, V^\beta,\label{8}
\end{equation}
where $C_{\mu\alpha\nu\beta}$ denotes the Weyl tensor. 

Equivalently, the electric part of the Weyl tensor may also be written as

\begin{eqnarray}
E_{\alpha \beta}&=&\mathcal{E}_1\left(K_\alpha L_\beta+L_\alpha K_\beta\right)
+\mathcal{E}_2\left(K_\alpha K_\beta-\frac{1}{3}h_{\alpha \beta}\right)\nonumber \\&+&\mathcal{E}_3\left(L_\alpha L_\beta-\frac{1}{3}h_{\alpha \beta}\right), \label{13}
\end{eqnarray}
where explicit expressions for the three scalars $\mathcal{E}_1$, $\mathcal{E}_2$, $\mathcal{E}_3$ are given in the Appendix \ref{B}.

Next, let us calculate the electric part of the Riemann tensor (the magnetic part vanishes identically), which is defined by
\begin{equation}
Y^\rho_\beta=V^\alpha V^\mu R^\rho_{\alpha \beta \mu}.
\label{29}
\end{equation}

After some lengthy calculations we find;

\begin{eqnarray}
Y_{\alpha \beta}&=&Y_{TF_1}\left(K_\alpha L_\beta+K_\beta L_\alpha\right)
+Y_{TF_2}\left(K_\alpha K_\beta-\frac{1}{3}h_{\alpha \beta}\right)\nonumber \\
&+&Y_{TF_3}\left(L_\alpha L_\beta-\frac{1}{3}h_{\alpha \beta}\right)+\frac{1}{3} Y_T h_{\alpha \beta},
\label{30}
\end{eqnarray}
where
\begin{equation}
Y_T=4\pi(\mu+3P),
\label{31}
\end{equation}

\begin{equation}
Y_{TF_1}=\mathcal{E}_1-4\pi \Pi_{KL},
\label{32}
\end{equation}
\begin{equation}
Y_{TF_2}=\mathcal{E}_2-4\pi \Pi_2,
\label{33}
\end{equation}

\begin{equation}
Y_{TF_3}=\mathcal{E}_3-4\pi \Pi_3.
\label{34}
\end{equation}
Finally, we shall find the tensor associated with the double dual of Riemann tensor, defined as
\begin{equation}
X_{\alpha \beta}=^*R^{*}_{\alpha \gamma \beta \delta}V^\gamma
V^\delta=\frac{1}{2}\eta_{\alpha\gamma}^{\quad \epsilon
\rho}R^{*}_{\epsilon \rho\beta\delta}V^\gamma V^\delta,
\label{35}
\end{equation}
with $R^*_{\alpha \beta \gamma \delta}=\frac{1}{2}\eta
_{\epsilon \rho \gamma \delta} R_{\alpha \beta}^{\quad \epsilon
\rho}$,
where $\eta_{\epsilon \rho \gamma \delta}$ denotes the permutation symbol.

Thus, we find
\begin{eqnarray}
X_{\alpha \beta}&=&X_{TF_1}\left(K_\alpha L_\beta+K_\beta L_\alpha\right)+
X_{TF_2}\left(K_\alpha K_\beta-\frac{1}{3}h_{\alpha \beta}\right)\nonumber \\
&+&X_{TF_3}\left(L_\alpha L_\beta-\frac{1}{3}h_{\alpha \beta}\right)+\frac{1}{3}X_T h_{\alpha \beta},
\label{36}
\end{eqnarray}
where
\begin{equation}
X_T=8\pi \mu,
\label{37}
\end{equation}

\begin{equation}
X_{TF_1}=-(\mathcal{E}_1+4\pi \Pi_{KL}),
\label{38}
\end{equation}
\begin{equation}
X_{TF_2}=-\left(\mathcal{E}_2+4\pi \Pi_2\right),
\label{39}
\end{equation}

\begin{equation}
X_{TF_3}=-\left(\mathcal{E}_3+4\pi \Pi_3\right).
\label{40}
\end{equation}

The  scalars $Y_T$, $Y_{TF1}$, $Y_{TF2}$,$ Y_{TF3}$, $X_T$, $X_{TF1}$, $X_{TF2}$, $X_{TF3}$, are the structure scalars for our system.

Next, we shall need two differential equations which relate the spatial derivatives of the  physical variables and the Weyl tensor, obtained  from Bianchi identities, they have been found before for the spherically symmetric and the cylindrically symmetric cases (see \cite{sc}, \cite{scc1} and references therein). For our case they  have been calculated in \cite{as}, and read

\begin{eqnarray}
&&\frac{{\cal E}_{1\theta}}{r}+\frac{1}{3}(2{\cal E}_2-{\cal E}_3)^\prime+\frac{{\cal E}_1}{r}\left(\frac{2B_\theta}{B}+
\frac{D_\theta}{D}\right)\nonumber \\&+&{\cal E}_2\left(\frac{B^\prime}{B}+\frac{D^\prime}{D}+\frac{1}{r}\right) -{\cal E}_3 \left(\frac {B^\prime}{B}+\frac{1}{r}\right)\nonumber \\&=&
\frac{4\pi}{3}\left(2\mu+3P\right)^\prime
+4\pi \left[\mu+P+\frac{1}{3}(2\Pi_2-\Pi_3)\right]\frac{A^\prime}{A}\nonumber \\&+&4\pi \Pi_{KL}\frac{A_\theta}{Ar},
\label{55}
\end{eqnarray}

\begin{eqnarray}
&&{\cal E}^{\prime}_1+\frac{1}{3r}(2{\cal E}_3-{\cal E}_2)_\theta+{\cal E}_1\left(\frac{2B^\prime}{B}+
\frac{D^\prime}{D}+\frac{2}{r}\right)-\frac{{\cal E}_2 B_\theta}{Br}\nonumber \\&+&\frac{{\cal E}_3}{r} \left(\frac{B_\theta}{B}+\frac{D_\theta}{D}\right)=
\frac{4\pi}{3r}\left(2\mu+3P \right)_\theta \nonumber \\&+&4\pi\left[\mu+P+\frac{1}{3}(2\Pi_{3}-\Pi_2)\frac{A_\theta}{Ar}\right]+4\pi \Pi_{KL}\frac{A^\prime}{A},
\label{56}
\end{eqnarray}

which, using (\ref{31})-(\ref{34}) and (\ref{37})-{\ref{40}), may be written in terms of structure scalars, producing

\begin{eqnarray}
\frac{8 \pi \mu^\prime}{3}&=&\frac{1}{r}\left[Y_{TF1\theta}+8\pi \Pi_{KL\theta} +(Y_{TF1}+8\pi \Pi_{KL})(\ln {B^2 D)}_{\theta}
\right]\nonumber \\&+&\left[\frac{2}{3}(Y_{TF2}^\prime +8\pi \Pi_2^\prime)+(Y_{TF2}+8\pi \Pi_2)(\ln{BDr})^{\prime}\right]\nonumber \\
&-&\left[\frac{1}{3}(Y_{TF3}^\prime+8\pi\Pi_3^\prime)+(Y_{TF3}+8\pi \Pi_3)(\ln{Br})^{\prime}\right],
\label{57}
\end{eqnarray}

\begin{eqnarray}
\frac{8 \pi \mu_{\theta}}{3r}&=&-\frac{1}{r}\left[\frac{1}{3}(Y_{TF2\theta}+8\pi \Pi_{2\theta})+(Y_{TF2}+8\pi \Pi_2)(\ln B)_{\theta}\right]\nonumber \\&+&\frac{1}{r}
\left[\frac{2}{3}(Y_{TF3\theta}+8\pi \Pi_{3\theta})+(Y_{TF3}+8\pi \Pi_3)(\ln{ BD})_{\theta}\right]\nonumber \\&+&
\left[Y_{TF1}^\prime+8\pi \Pi_{KL}^\prime+(Y_{TF1}+8\pi \Pi_{KL})(\ln{B^2 D r^2})^{\prime}\right],\nonumber
\label{58}
\end{eqnarray}

where prime  and subscript $\theta$ denote derivatives with respect to $r$  and $\theta$ respectively.

We have now available all the elements necessary to identify the complexity factors for  the fluid distribution under consideration. For doing so we recall our basic ansatz consisting in assuming that the simplest possible fluid (or at least one of them) is the  incompressible (constant energy density) fluid with isotropic pressure.

 Now, in \cite{as} it has been shown that  the necessary and sufficient conditions for the vanishing of the (invariantly defined) spatial derivatives of  the energy density  are $X_{TF1}=X_{TF2}=X_{TF3}=0$. In other words
\begin{equation}
X_{TF1}=X_{TF2}=X_{TF3}=0\Leftrightarrow  \mu^\prime=\mu_\theta=0.
\label{70n}\end{equation}

Therefore the homogeneous energy--density condition implies $X_{TF1}=X_{TF2}=X_{TF3}=0$, which in turn produces

\begin{equation}
Y_{TF1}=-8\pi \Pi_{KL};\quad Y_{TF2}=-8\pi \Pi_2;\quad Y_{TF3}=-8\pi \Pi_3.
\label{71n}\end{equation}

From the above it follows that the isotropic pressure condition would imply $Y_{TF1}= Y_{TF2}= Y_{TF3}=0$.

In other words, following the rationale exposed in the spherically symmetric case, it is reasonable to identify the three structure scalars $Y_{TF}$ (more precisely, their absolute values) with  the complexity factors. As in the previous  case, we notice that  they vanish for the incompressible (constant energy density) fluid with isotropic pressure, but may also vanish for inhomogeneous, anisotropic fluids, provided these two factors combine in such a way that they cancel the three complexity factors.

In the next subsections, just  to illustrate the way by means of which such models may be obtained,  we shall present  two solutions with vanishing complexity factors.

\subsection{The incompressible, isotropic  spheroid}

The first solution we shall present corresponds to the case where the complexity factors vanish because the energy density is homogeneous and the pressure is isotropic. This solution was previously obtained and analyzed in \cite{as}. Here we just present  it without details.  

Let us first notice that,  from (\ref{70n}), (\ref{38})--(\ref{40}) and $P_{xx}=P_{yy}=P_{zz}=P$, $P_{xy}=0$, $\mu=\mu_0=constant$, it follows  that such a solution is also conformally flat.

Next, for simplicity we shall assume the boundary surface $\Sigma$ to be defined by the equation:
\begin{equation}
r=r_1=constant,
\label{50ns}
\end{equation}
which is not the most general form of a possible  boundary surface.

Then, from  the above and  (\ref{25}) and (\ref{26}) it follows that
\begin{equation}
P\stackrel{\Sigma}{=}0,
\label{51ns}
\end{equation}
where $\stackrel{\Sigma}{=}$ means that both sides of the equation
are evaluated on $\Sigma$

Under the conditions above, (\ref{22}) and (\ref{23}) can be integrated to obtain:
\begin{equation}
P+\mu_0=\frac{\zeta}{A},
\label{52a}
\end{equation}
and
\begin{equation}
P+\mu_0=\frac{\xi(r)}{A},
\label{53a}
\end{equation}
where $\xi$ is  an arbitrary function of its argument.
Using boundary conditions (\ref{51ns}) in (\ref{52a}) (\ref{53a}) it follows that:
\begin{equation}
 A(r_1,\theta)=const.=\frac{\alpha}{\mu_0}, \qquad \zeta=constant.
\label{54}
\end{equation}
Finally, the metric for this model  can be written as follows

\begin{widetext}
 \begin{equation}
 ds^2=\frac{1}{(\gamma r^2+\delta +b r\cos \theta)^2}\left [-(\alpha r^2+\beta+a r\cos\theta)^2dt^2+dr^2+r^2d\theta ^2+r^2\sin ^2 \theta d\phi ^2\right ], \label{cfifm}
\end{equation}
\end{widetext}

from which,  the physical variables can be easily calculated, producing

 \begin{equation}
8\pi \mu = 12 \gamma \delta-3b^2,
\label{den1}
\end{equation}

\begin{equation}
8\pi P =(3b^2-12\gamma\delta )\left [1-\frac{\alpha
r_1^2+\beta}{\gamma r_1^2+\delta} \frac{\gamma
r^2+\delta+br\cos\theta}{\alpha r^2+\beta +a
r\cos\theta}\right ],\label{epf}
\end{equation}
where  $b, \gamma, \delta$ are constants, and 

\begin{equation}
\zeta=\mu_0\frac{\alpha r_1^2+\beta}{\gamma
r_1^2+\delta},\quad  a=\frac{\alpha r_1^2+\beta}{\gamma
r_1^2+\delta}b,
\label{jcn}
\end{equation}
in order to satisfy  the junction condition  (\ref{51ns}).

It is important to stress the fact that this solution cannot be matched to any Weyl exterior, except in the spherically symmetric case, even though it  has a surface of vanishing pressure (see \cite{as} for details). As shown in \cite{as} this  result is a consequence of  the energy density homogeneity and the pressure isotropy. So, to find matchable solutions we should relax these  two conditions. Also, it is worth mentioning that  the above  result is in agreement with  previous works  indicating that static, perfect fluid (isotropic in pressure) sources are spherical (see \cite{prueba} and references therein).

\subsection{Anisotropic inhomogeneous spheroids}

\noindent  In order to obtain a metric smoothly matchable to any Weyl space--time  we shall consider  a solution with vanishing complexity factors but inhomogeneous energy density and anisotropic pressure (see details in \cite{cax})

The metric variables of the solution are
\begin{eqnarray}
  A(r,\theta) &=& \frac{a_1 r \sin \theta}{b_1 r^2+b_2},\label{ns2a} \\
  B(r,\theta)&=& \frac{1}{b_1 r^2+b_2}, \\
  D(r,\theta) &=& \frac{b_1 r^2-b_2}{b_1 r^2+b_2} F\left(\frac{r \cos \theta}{b_1 r^2-b_2}\right). \label{ns1a}
\end{eqnarray}
It is a simple matter to check that  the vanishing complexity factors conditions (\ref{YTF1})--(\ref{YTF3}) in Appendix \ref{C}, are satisfied for (\ref{ns2a})--(\ref{ns1a}).

From the above, using the Einstein equations (\ref{24})--(\ref{26}), one obtains  for the physical variables.

\begin{eqnarray}
  8\pi \mu = 12b_1 b_2\nonumber \\-\frac{(b_1 r^2+b_2)^2}{(b_1 r^2-b_2)^2}\left[\frac{4b_1 b_2  r^2 cos^2\theta}{(b_1 r^2-b_2)^2}+1\right]\frac{F_{zz}}{F}, \\
   8\pi P= -12b_1 b_2\nonumber \\+\frac{(b_1 r^2+b_2)^2}{3(b_1 r^2-b_2)^2}\left[\frac{4 b_1 b_2  r^2 cos^2\theta}{(b_1 r^2-b_2)^2} +1\right]\frac{F_{zz}}{F} ,\\
  8\pi \Pi_2\equiv  8\pi (P_{xx}-P_{zz}) = \frac{F_{zz}}{4F}\frac{(b_1 r^2+b_2)^2}{(b_1 r^2-b_2)^2} \sin^2 \theta,\\
  8\pi \Pi_3\equiv  8\pi (P_{yy}-P_{zz}) = \frac{F_{zz}}{4F}\frac{(b_1 r^2+b_2)^4 cos^2\theta}{(b_1 r^2-b_2)^4},\\
   8\pi \Pi_{KL}\equiv 8\pi P_{xy}=- \frac{F_{zz}}{2F}\frac{r(b_1 r^2+b_2)^3}{(b_1 r^2-b_2)^3} \sin 2\theta,
\end{eqnarray}
where $a_1, b_1, b_2$ are constant, and 
\begin{equation}
F(z)\equiv F\left(\frac{r \cos\theta}{b_1 r^2-b_2}\right).
\end{equation}

It is not difficult  to find a range of values of the parameters, for which the physical behavior of physical variables is acceptable and the metric may be matched smoothly on the  boundary surface to a Weyl solution.

\subsection*{The static hyperbolically symmetric case}
Motivated by a new version of the Schwarzschild black hole proposed in  \cite{1w, 2w}, there has been a renewed interest in self--gravitating systems admitting hyperbolical symmetry. 

In this picture, the  space--time outside the horizon is represented by the usual Schwarzschild metric, whereas  the region inner to the horizon  is described  by the line element 
\begin{eqnarray}
ds^2&=&\left(\frac{2M}{R}-1\right)dt^2-\frac{dr^2}{\left(\frac{2M}{R}-1\right)}-R^2d\Omega^2, \nonumber \\ d\Omega^2&=&d\theta^2+\sinh^2 \theta d\phi^2,
\label{w3}
\end{eqnarray}

which is a static solution with the $(\theta  ,\phi )$ space describing a positive Gaussian curvature, admitting the four Killing vectors
\begin{eqnarray}
\mathbf{\chi }_{(\mathbf{0})} = \partial _{\mathbf{t}}, \quad {\bf \chi_{(2)}}=-\cos \phi \partial_{\theta}+\coth\theta \sin\phi \partial_{\phi}\nonumber \\
{\bf \chi_{(1)}}=\partial_{\phi} \quad {\bf \chi_{(3)}}=\sin \phi \partial_{\theta}+\coth\theta \cos\phi \partial_{\phi}.
\label{2cmhy}
\end{eqnarray}

A solution to the Einstein equations of the form given by (\ref{w3}), defined by the hyperbolic symmetry  (\ref{2cmhy}), was first considered by Harrison \cite{Ha}, and has been more recently the subject of research  in different contexts (see \cite{ Ga, Ri, mim, Ka, Ma, mimII} and references therein).

Our purpose here is to present some exact solutions to Einstein equations endowed with the symmetry given by (\ref{2cmhy}), satisfying the vanishing  complexity factor condition and which might  serve as the source of (\ref{w3}), see \cite{rg4} for details.

Thus, let us consider hyperbolically  symmetric distributions of  static
fluid, which for the sake of completeness we assume to be locally anisotropic and  which may be (or may be not) bounded from the exterior by a
surface $\Sigma^{e}$ whose equation is $r=r_{\Sigma^{e}}=\rm constant$. On the other hand as it appears from the study of such fluids (see \cite{2w} for details) the fluid distribution cannot  fill the central region, in which case we may assume that such a region is represented by an empty vacuole, implying that the fluid distribution is also bounded from the inside  by a
surface $\Sigma^{i}$ whose equation is $r=r_{\Sigma^{i}}=\rm constant$.

\noindent
The line element is given in polar  coordinates  by

\begin{equation}
ds^2=e^{\nu} dt^2 - e^{\lambda} dr^2 -
r^2 \left( d\theta^2 + \sinh^2\theta d\phi^2 \right),
\label{metric}
\end{equation}

\noindent
where, due to the imposed symmetry, $\nu(r)$ and $\lambda(r)$ are exclusively functions of  $r$. We
number the coordinates: $x^0=t; \, x^1=r; \, x^2=\theta; \, x^3=\phi$.

\noindent
The metric (\ref{metric}) has to satisfy Einstein field equations

\begin{equation}
G^\nu_\mu=8\pi T^\nu_\mu.
\label{Efeq}
\end{equation}
\noindent

We may write for the
 energy momentum tensor
\begin{eqnarray}
{T}_{\alpha\beta}&=& (\mu+P) V_\alpha V_\beta-P g _{\alpha \beta} +\Pi_{\alpha \beta},
\label{6bis}
\end{eqnarray}
or
\begin{equation}
{T}_{\alpha\beta}= (\mu+P_{zz}) V_\alpha V_\beta-P_{zz} g _{\alpha \beta} +(P_{xx}-P_{zz}) K_\alpha  K_\beta.
\label{6}
\end{equation}

Since we choose the fluid to be comoving in our coordinates, then
\begin{equation}
V^\alpha =(e^{-\nu/2}, 0, 0, 0); \quad  V_\alpha=(e^{\nu/2}, 0, 0, 0),
\label{m1}
\end{equation}
and

\begin{equation}
 K_{\alpha}=(0, -e^{\lambda/2}, 0, 0),
\label{7}
\end{equation}

It would be useful  to express the anisotropic tensor   in the form

\begin{eqnarray}
\Pi_{\alpha \beta}=\Pi\left(K_\alpha K_\beta+\frac{h_{\alpha
\beta}}{3}\right) \label{6bb},
\end{eqnarray}

with $h_{\mu \nu}=g_{\mu\nu}-V_\nu V_\mu$,
\begin{eqnarray}
\Pi=P_{xx}-P_{zz},
\label{6bb}
\end{eqnarray}

and
\begin{equation}
P=\frac{P_{xx}+2P_{zz}}{3}.
\label{7P}
\end{equation}

Since the Lie derivative and the partial derivative commute, then
\begin{equation}
\mathcal{L}_\chi G_{\alpha \beta}=8\pi \mathcal{L}_\chi T_{\alpha \beta}=0,
\label{ccm1}
\end{equation}
for any $\chi$ defined by  (\ref{2cmhy}), implying that all physical variables only depend on $r$.

If  the fluid is bounded from the exterior by a hypersurface $\Sigma^e$ described by the equation  $r=r_{\Sigma^e}=constant$, then the smooth matching of (\ref{w3}) and (\ref{metric}) on $\Sigma^e$ requires the fulfillment of the Darmois conditions, imposing the continuity of the first and the second fundamental forms, which imply

\begin{equation}
e^{\nu_{\Sigma^e}}=\frac{2M}{r_{\Sigma^e}}-1, \qquad e^{\lambda_{\Sigma^e}}=\frac{1}{\frac{2M}{r_{\Sigma^e}}-1},\qquad P_{xx}(r_{\Sigma^e})=0,
\label{j1}
\end{equation}
and the continuity of the mass function $m(r)$ defined below. 
If we assume that the central region is surrounded by an empty cavity whose delimiting surface is $r=r_{\Sigma^i}=constant$, then the fulfillment of Darmois conditions on $\Sigma^i$ implies
\begin{equation}
e^{\nu_{\Sigma^i}}=1, \qquad e^{\lambda_{\Sigma^i}}=1,\qquad P_{xx}(r_{\Sigma^i})=0,
\label{j2}
\end{equation}
and $m(r_{\Sigma^i})=0$.

\noindent The non--vanishing components of the Einstein equations for the metric  (\ref{metric}) and the energy momentum tensor  (\ref{6}) are

\begin{eqnarray}
  8\pi \mu &=& -\frac{(e^{-\lambda}+1)}{r^2}+\frac{\lambda ^\prime}{r}e^{-\lambda},\label{mu} \\
  8\pi P_r &=&\frac{(e^{-\lambda}+1)}{r^2}+\frac{\nu ^\prime}{r}e^{-\lambda}, \label{pr}\\
  8\pi P_\bot&=& \frac{e^{-\lambda}}{2}\left (\nu^{\prime\prime}+\frac{{\nu^\prime}^2}{2}-\frac{\lambda^\prime \nu^\prime}{2}+\frac{\nu^\prime}{r}-\frac{\lambda^\prime}{r}\right),\label{pbot}
\end{eqnarray}
where we have used the standard notation $P_{xx}\equiv P_r$ and $P_{zz}=P_{yy}\equiv P_\bot$, and primes denote derivatives with respect to $r$.

It is worth stressing the differences between these equations and the corresponding to the spherically symmetric case.

From the equations above or using the conservation laws $T^\alpha_{\beta;\alpha}=0$ we obtain, besides the identity $\dot\mu = 0$ (where dot denotes derivative with respect to $t$), the corresponding hydrostatic  equilibrium equation (the generalized Tolman--Oppenheimer--Volkoff equation)
\begin{eqnarray}
  P_r^\prime+(\mu+P_r)\frac{\nu^\prime}{2}+\frac{2}{r} \Pi&=& 0\label{CPx}.
\end{eqnarray}

Let us now define the mass function $m=m(r)$.
For doing so, let us notice that using (\ref{w3}) we have that  outside the fluid distribution (but inside the horizon)
\begin{equation}
M=-\left(\frac{R}{2} \right)R^3_{232},
\label{m1n}
\end{equation}
where the Riemann tensor component $R^3_{232}$, has been calculated with (\ref{w3}).

Then generalizing the above definition of mass  for the interior of the fluid distribution we may write
\begin{equation}
m(r)=-\left(\frac{r }{2}\right)R^3_{232}=\frac{r (1+e^{-\lambda})}{2}
\label{m3n}
\end{equation}
where now the Riemann tensor component is calculated with (\ref{metric}).

Feeding back (\ref{m3n}) into (\ref{mu}) we obtain

\begin{equation}
m^\prime(r)=-4\pi r^2 \mu \Rightarrow m=-4\pi \int^r_0{\mu r^2dr}.
\label{m2}
\end{equation}

Since $m$ as defined by (\ref{m3n}) is a positive quantity,  then $\mu$ should be  negative  and therefore the weak energy condition is violated, a result already obtained in \cite{mimII}. However it is important to stress that our definition of mass function differs from the one introduced in \cite{mimII}. In particular our $m$ is positive defined whereas the expression used in \cite{mimII} is negative (for the hyperbolically symmetric fluid). 

 The following   comments are  in order at this point.  
 \begin{itemize}
\item If the energy density is regular everywhere then  the mass function must vanish at the center as $m\sim r^3$,  this implies (as it follows from (\ref{m3n})) that the fluid cannot fill the space in the neighborhood of  the center, i.e. there is a cavity around the center which may be, either  empty,  or  filled with a fluid distribution non endowed with hyperbolical symmetry. Thus the hyperbolically symmetric fluid spans from a minimal value of the coordinate $r$ until its external boundary. For the extreme case $\mu=constant$, this minimal value $r_{min.}$ is defined by $-\frac{8\pi}{3} \mu r^2_{min.}>1$. Obviously, if the  energy density is singular in the neighborhood of the center, then this region must also be excluded by physical reasons.
\end{itemize}

From the  above it follows that,  strictly speaking,  we should write instead of (\ref{m2})

\begin{equation}
 m=4\pi \int^r_{r_{min}}{\vert \mu \vert r^2dr},
\label{m3}
\end{equation}
where due to the fact that $\mu$ is negative, we have replaced it by $-\vert \mu \vert$ (as we shall do from now on).

The situation described above is fully consistent with the results obtained in \cite{2w} where it was shown that test particles cannot reach the center for any finite value of its energy.

Next, using (\ref{pr}) and (\ref{m3n}) we obtain
\begin{equation}
\nu^\prime=2\frac{4\pi r^3 P_r-m}{r(2 m-r)},
\label{m3}
\end{equation}
from which we may write (\ref{CPx}) as

\begin{eqnarray}
  P_r^\prime+(P_r-\vert \mu\vert)\frac{4\pi r^3 P_r-m}{r(2 m-r)}+\frac{2}{r} \Pi&=& 0\label{m4}.
\end{eqnarray}

This is the hydrostatic equilibrium equation for our fluid. Let us analyze in some detail  the physical meaning of its different terms. 

The first term is just the gradient of pressure, which is usually negative  and opposing gravity. The second term describes the gravitational ``force'' and contains two different contributions: on the one hand the term $P_r-\vert \mu \vert$ which we expect to be negative (or zero for the stiff equation of state) and is usually interpreted as the ``passive gravitational mass density'' (p.g.m.d.), and on the other hand  the term $4\pi r^3 P_r-m$  that is proportional to the ``active gravitational mass''  (a.g.m.), and which  is negative if $4\pi r^3 P_r<m$. Finally the third term describes the effect of the pressure anisotropy, whose sign depends on the difference between principal stresses.
Two important remarks are in order at this point:
\begin{itemize}
\item It is worth stressing that while the self--regenerative pressure effect (described by the  $4\pi r^3 P_r$ term in (\ref{m4})) has the same sign  as in the spherically symmetric case, the mass function contribution in the second term has the opposite sign with respect to the latter case. This of course is due to the fact that the energy density is negative.
\item If, both, the p.g.m.d. and the a.g.m. are negative, the final effect of the gravitational interaction would be as usual, to oppose the negative pressure gradient. However, because of the equivalence principle, a negative  p.g.m.d. implies a negative inertial mass, which in turn implies that the hydrostatic force term (the pressure gradient and the anisotropic term), and the gravitational force term, switch their roles with respect to the positive energy density  case.
\end{itemize}

In this case  the complexity factor  $Y_{TF}$ is given by 

\begin{equation}
Y_{TF}= 4\pi \Pi+{\cal E},
\label{defYTF}
\end{equation}
or  
\begin{equation}
Y_{TF}=8\pi \Pi+\frac{4\pi}{r^3} \int^r_0{\tilde r^3 \vert \mu \vert' d\tilde r},
\label{defYTFbis}
\end{equation}
where ${\cal E}$ is the scalar defining the electric part of the Weyl tensor (the magnetic part vanishes identically as in the spherically symmetric case).

Also, as in the spherically  symetric case $Y_{TF}$ encompasses the influence of the local anisotropy of pressure and density inhomogeneity on the Tolman mass. Or, in other words, $Y_{TF}$ describes how these two factors modify the value of the Tolman mass, with respect to its value for the homogeneous isotropic fluid. 

Indeed, in this case it can be shown that  the following expressions for the Tolman  mass can be obtained (see \cite{rg4} for details)
\begin{equation}\label{masaT1}
  m_T=\frac{(cosh\pi-1)}{4}e^{(\nu-\lambda)/2}r^2\nu^\prime,
\end{equation}
and 
\begin{eqnarray}
m_T  &=&  (m_T)_{\Sigma^e} \left(\frac{r}{r_{\Sigma^e}}\right)^3\nonumber \\
 &+& \frac{(cosh\pi-1)}{2} r^3 \int^{r_{\Sigma^e}}_r{\frac{e^{(\nu+\lambda)/2}}{\tilde r} Y_{TF}d\tilde r}.
\label{emtebis}
\end{eqnarray}

We shall next present two models endowed with hyperbolical symmetry and satisfying the vanishing complexity factor condition.

\subsubsection{A model with vanishing complexity factor and vanishing radial pressure}
Since the vanishing complexity factor is not enough to close the full system of Einstein equations we have to impose an additional restriction in order to obtain a specific model. Here we shall assume (besides the vanishing complexity factor), the condition $P_r=0$.

\noindent Thus, assuming $P_r=0$, we obtain from  (\ref{pr})
\begin{equation}\label{nu1}
  \nu^\prime=-\frac{2g}{(2g-1)r},
\end{equation}
\noindent where $g$ is defined by
\begin{equation}\label{g2}
  e^{-\lambda}=2g-1.
\end{equation}

\noindent Next, imposing  $Y_{TF}=0$ in  (\ref{emtebis}) it follows that

\begin{equation}\label{ytf0}
  m_T=(m_T)_{\Sigma^e} \frac{r^3}{r^3_\Sigma}.
\end{equation}

\noindent The combination of (\ref{masaT1}), (\ref{nu1}), (\ref{g2}) and  (\ref{ytf0})  produces

\begin{equation}\label{nuytf}
  e^\nu=\frac{4(m^2_T)_{\Sigma^e} r^4}{r^6_\Sigma (cosh\pi-1)^2}\frac{(2g-1)}{g^2}.
\end{equation}

On the other hand the condition $Y_{TF}=0$  may be written as
\begin{equation}
g^\prime r (1-g)+g(5g-2)=0,
\label{sol1}
\end{equation}

\noindent whose solution reads
\begin{equation}
C_2r^{10}=\frac{g^5}{(5g-2)^3}\label{g3},
\end{equation}
where $C_2$ is a constant of integration.

\noindent Then for the physical variables we obtain

\begin{eqnarray}
 \vert \mu \vert  &=& \frac{3}{4\pi r^2}\frac{g(2g-1)}{(g-1)}, \\
  P_\bot &=& \frac{3}{8\pi r^2}\frac{g^2}{(g-1)}.
\end{eqnarray}

In this case,  the fluid distribution  is restricted by a minimal value of the $r$ coordinate, satisfying $g(r_{min})>1$. The specific value of $r_{min}$ is obtained from (\ref{g3}). For $0<r<r_{min}$ we may assume, as in precedent models,  an empty cavity surrounding the center. Also as in precedent models, the discontinuity of the mass function across $\Sigma^i$ implies that a thin shell appears on it.  Finally, since the radial pressure is assumed to be zero, both the a.g.m.  and the p.g.m.d. are negative.

\subsubsection{A  model with the stiff equation of state and vanishing complexity factor}
Finally we shall consider a solution satisfying the so called stiff equation of state  proposed by Zeldovich \cite{zh} and which is thought to be suitable to describe ultradense matter, and the vanishing complexity factor. 

 In its original form the stiff equation of state assumes that energy density equals pressure (in relativistic units). In our case we shall assume
\begin{equation}
\vert \mu \vert=P_r.
\label{s1}
\end{equation}
then (\ref{m4}) becomes
\begin{eqnarray}
  P_r^\prime+\frac{2}{r} \Pi&=& 0\label{m4z}.
\end{eqnarray}

Using this latter condition in (\ref{defYTFbis}) with $Y_{TF}=0$, and feeding back the resulting expression into (\ref{m4z}) one obtains

\begin{equation}
P^{\prime \prime}_r+\frac{3}{r}P^\prime_r=0,
\label{z8}
\end{equation}
whose solution reads
\begin{equation}
P_r=\frac{b}{r^2}-a,
\label{z9}
\end{equation}
where $a$ and $b$ are two positive constants of integration.

Then from (\ref{m3n}) and (\ref{m2}) it follows at once
\begin{equation}
m=4\pi r\left(b-\frac{ar^2}{3}\right),
\label{z10}
\end{equation}
from which we easily obtain $\lambda$. Finally, feeding back these expressions into (\ref{m3}) we may obtain $\nu$.

Assuming the fluid distribution to be bounded from the exterior by the surface $\Sigma^e$ described by $r=r_{\Sigma^e}=constant$, then we may write
\begin{equation}
P_r=b\left(\frac{1}{r^2}-\frac{1}{{r^2_{\Sigma^e}}}\right).
\label{z11}
\end{equation}

So far we have only considered static configurations. In the next section we shall tackle the problem of defining complexity for dynamic dissipative fluids.

\section{Dynamic spherically symmetric fluids}
As mentioned before, when dealing with non--static dissipative fluids we encounter two additional problems  to define complexity. On the one hand we have to include the dissipative flux in the variable measuring the complexity of the structure. On the other hand we have also to describe the complexity of the pattern of evolution.

In other words, in the dynamic case we have to ask ourselves two questions instead of one, namely, what is the simplest fluid? and, what is the simplest mode of evolution of the fluid? 
These two questions, although related in some way, are completely different. A ``simple'' fluid may evolve exhibiting a very complex pattern of evolution, while a fluid with a high degree of complexity may evolve through a simple pattern of evolution.

This problem was studied in detail in \cite{ch2}, here we shall sketch  the main  steps leading to the appropriate definition of complexity for dynamic fluids and discuss about the answer to the question, what is the simplest mode of evolution of the fluid? Afterwards we shall present some solutions.

We shall restrict the discussion to the spherically symmetric case.

\subsection{The dynamic spherically symmetric case}
So, let us  consider a spherically symmetric distribution  of collapsing
fluid, which may be  bounded by a spherical surface $\Sigma$, or not. The fluid is
assumed to be locally anisotropic (principal stresses unequal) and undergoing dissipation in the
form of heat flow (diffusion approximation). 

Choosing comoving coordinates, the general
interior metric can be written
\begin{equation}
ds^2=-A^2dt^2+B^2dr^2+R^2(d\theta^2+\sin^2\theta d\phi^2),
\label{1}
\end{equation}
where $A$, $B$ and $R$ are functions of $t$ and $r$ and are assumed
positive. We number the coordinates $x^0=t$, $x^1=r$, $x^2=\theta$
and $x^3=\phi$. Observe that $A$ and $B$ are dimensionless, whereas $R$ has the same dimension as $r$.

The energy--momentum tensor $T_{\alpha\beta}$ of the fluid distribution
has the form
\begin{eqnarray}
T_{\alpha\beta}&=&(\mu +
P_{\perp})V_{\alpha}V_{\beta}+P_{\perp}g_{\alpha\beta}+(P_r-P_{\perp})\chi_{
\alpha}\chi_{\beta}\nonumber \\&+&q_{\alpha}V_{\beta}+V_{\alpha}q_{\beta}
, \label{3}
\end{eqnarray}
where $\mu$ is the energy density, $P_r$ the radial pressure,
$P_{\perp}$ the tangential pressure, $q^{\alpha}$ the heat flux, $V^{\alpha}$ the four velocity of the fluid,
and $\chi^{\alpha}$ a unit four vector along the radial direction. These quantities
satisfy
\begin{eqnarray}
V^{\alpha}V_{\alpha}=-1, \;\; V^{\alpha}q_{\alpha}=0, \;\; \chi^{\alpha}\chi_{\alpha}=1,\;\;
\chi^{\alpha}V_{\alpha}=0.
\end{eqnarray}

Or in the equivalent (canonical) form
\begin{equation}
T_{\alpha \beta} = {\mu} V_\alpha V_\beta + P h_{\alpha \beta} + \Pi_{\alpha \beta} +
q \left(V_\alpha \chi_\beta + \chi_\alpha V_\beta\right) \label{Tab}
\end{equation}
with
$$ P=\frac{P_{r}+2P_{\bot}}{3}, \qquad h_{\alpha \beta}=g_{\alpha \beta}+V_\alpha V_\beta,$$

$$\Pi_{\alpha \beta}=\Pi\left(\chi_\alpha \chi_\beta - \frac{1}{3} h_{\alpha \beta}\right), \qquad \Pi=P_{r}-P_{\bot},$$ where $q$ is a function of $t$ and $r$.

Since we are considering comoving observers, we have
\begin{eqnarray}
V^{\alpha}&=&A^{-1}\delta_0^{\alpha}, \;\;
q^{\alpha}=qB^{-1}\delta^{\alpha}_1, \;\;
\chi^{\alpha}=B^{-1}\delta^{\alpha}_1, 
\end{eqnarray}

It is worth noticing that, both, bulk and shear viscosity could be easily introduced to the system through a redefinition of  the radial and tangential pressures, $P_r$ and
$P_{\perp}$. Also,   dissipation in the free streaming approximation could be introduced by redefining  $\mu,  P_r$ and $q$. 

The Einstein equations for (\ref{1}) and (\ref{Tab}), are explicitly written  in Appendix \ref{D}.

The acceleration $a_{\alpha}$ and the expansion $\Theta$ of the fluid are
given by
\begin{equation}
a_{\alpha}=V_{\alpha ;\beta}V^{\beta}, \;\;
\Theta={V^{\alpha}}_{;\alpha}. \label{4b}
\end{equation}
and its  shear $\sigma_{\alpha\beta}$ by
\begin{equation}
\sigma_{\alpha\beta}=V_{(\alpha
;\beta)}+a_{(\alpha}V_{\beta)}-\frac{1}{3}\Theta h_{\alpha\beta},
\label{4a}
\end{equation}
from which we easily obtain 

\begin{equation}
a_1=\frac{A^{\prime}}{A}, \;\; a=\sqrt{a^{\alpha}a_{\alpha}}=\frac{A^{\prime}}{AB}, \label{5c}
\end{equation}

\begin{equation}
\Theta=\frac{1}{A}\left(\frac{\dot{B}}{B}+2\frac{\dot{R}}{R}\right),
\label{5c1}
\end{equation}

\begin{equation}
\sigma_{11}=\frac{2}{3}B^2\sigma, \;\;
\sigma_{22}=\frac{\sigma_{33}}{\sin^2\theta}=-\frac{1}{3}R^2\sigma,
 \label{5a}
\end{equation}
where
\begin{equation}
\sigma^{\alpha\beta}\sigma_{\alpha\beta}=\frac{2}{3}\sigma^2,
\label{5b}
\end{equation}
with
\begin{equation}
\sigma=\frac{1}{A}\left(\frac{\dot{B}}{B}-\frac{\dot{R}}{R}\right),\label{5b1}
\end{equation}
where the  prime stands for $r$
differentiation and the dot stands for differentiation with respect to $t$.

Next, the mass function $m(t,r)$ introduced by Misner and Sharp
\cite{Misner}  reads
\begin{equation}
m=\frac{R^3}{2}{R_{23}}^{23}
=\frac{R}{2}\left[\left(\frac{\dot R}{A}\right)^2-\left(\frac{R^{\prime}}{B}\right)^2+1\right],
 \label{17masa}
\end{equation}

and introducing the proper time derivative $D_T$
given by
\begin{equation}
D_T=\frac{1}{A}\frac{\partial}{\partial t}, \label{16}
\end{equation}
we can define the velocity $U$ of the collapsing
fluid  as the variation of the areal radius with respect to proper time, i.e.
\begin{equation}
U=D_TR, \label{19}
\end{equation}
where $R$ defines the areal radius of a spherical surface inside the fluid distribution (as
measured from its area).

Then (\ref{17masa}) can be rewritten as
\begin{equation}
E \equiv \frac{R^{\prime}}{B}=\left(1+U^2-\frac{2m}{R}\right)^{1/2}.
\label{20x}
\end{equation}
Using (\ref{20x}) we can express (\ref{17a}) as
\begin{equation}
4\pi q=E\left[\frac{1}{3}D_R(\Theta-\sigma)
-\frac{\sigma}{R}\right],\label{21a}
\end{equation}
where   $D_R$ denotes the proper radial derivative,
\begin{equation}
D_R=\frac{1}{R^{\prime}}\frac{\partial}{\partial r}.\label{23a}
\end{equation}
Using (\ref{12})-(\ref{14}) with  (\ref{23a}) we obtain from
(\ref{17masa})

\begin{eqnarray}
D_Rm=4\pi\left(\mu+q\frac{U}{E}\right)R^2,
\label{27Dr}
\end{eqnarray}
which implies
\begin{equation}
m=4\pi\int^{r}_{0}\left( \mu +q\frac{U}{E}\right)R^2R^\prime dr, \label{27intcopy}
\end{equation}
satisfying the regular condition  $m(t,0)=0$.

Integrating (\ref{27intcopy}) we find
\begin{equation}
\frac{3m}{R^3} = 4\pi {\mu} - \frac{4\pi}{R^3} \int^r_0{R^3\left(D_R{ \mu}-3 q \frac{U}{RE}\right) R^\prime dr}.
\label{3m/R3}
\end{equation}
\\

\subsection{Defining complexity for the dynamic fluid}
As we have already mentioned, in the dynamic case the definition of a quantity measuring the complexity of the system poses two  additional problems with respect to the static case. 

On the one hand, the definition of the complexity of the structure of the fluid, which in this case also involves  dissipative variables, and on the other hand the problem of defining    the complexity of the pattern of evolution of the system.

 For the static fluid distribution it was assumed  that  the scalar function $Y_{TF}$ is an appropriate measure of the complexity of the fluid, and therefore was identified as  the complexity factor. 
 
 We shall assume in the dynamic case  that $Y_{TF}$ still measures the complexity of the system, in what corresponds to the structure of the object, and we shall adopt initially an assumption about the simplest possible pattern of evolution. Specifically, we shall assume that the simplest   evolution pattern (one of them at least) is described by the homologous evolution. However, as we shall see below, this last condition might be  too stringent,  ruling out  many interesting scenarios from the astrophysical point of view and therefore we shall consider also other possible (less restrictive) mode of evolution which also could be used to describe the simplest mode of evolution, and which we call quasi--homologous.
 
 In order to provide the necessary mathematical expressions for carrying out our task, let us start by finding the expression for the Weyl tensor.
 
As is well known, in  the spherically symmetric case the Weyl tensor  ($C^{\rho}_{\alpha
\beta
\mu}$) is   defined by its ``electric'' part 
 $E_{\gamma \nu }$ alone, since its  ``magnetic'' part 
vanishes, with
  \begin{equation}
E_{\alpha \beta} = C_{\alpha \mu \beta \nu} V^\mu V^\nu,
\label{elec}
\end{equation}
where the electric part of Weyl tensor  may also be written as
\begin{equation}
E_{\alpha \beta}={\cal E} (\chi_\alpha \chi_\beta-\frac{1}{3}h_{\alpha \beta}).
\label{52}
\end{equation}
with
\begin{widetext}
\begin{eqnarray}
{\cal E}&=& \frac{1}{2 A^2}\left[\frac{\ddot R}{R} - \frac{\ddot B}{B} - \left(\frac{\dot R}{R} - \frac{\dot B}{B}\right)\left(\frac{\dot A}{A} + \frac{\dot R}{R}\right)\right]+ \frac{1}{2 B^2} \left[\frac{A^{\prime\prime}}{A} - \frac{R^{\prime\prime}}{R} + \left(\frac{B^{\prime}}{B} + \frac{R^{\prime}}{R}\right)\left(\frac{R^{\prime}}{R}-\frac{A^{\prime}}{A}\right)\right] - \frac{1}{2 R^2}.
\label{Ea}
\end{eqnarray}
\end{widetext}

Then, proceeding as in the previous cases (see \cite{ch2} for details) we obtain
\begin{eqnarray}
Y_T=4\pi(\mu+3 P_r-2\Pi) , \qquad
Y_{TF}={\cal E}-4\pi \Pi .\label{EY}
\label{EX}
\end{eqnarray}
  
Next, using  (\ref{12}), (\ref{14}), (\ref{15}) with (\ref{17masa}) and (\ref{Ea}) we obtain
\begin{equation}
\frac{3m}{R^3}=4\pi \left({\mu}-\Pi \right) - \cal{E},
\label{mE}
\end{equation}
which combined with (\ref{3m/R3})  and (\ref{EY}) produces

\begin{equation}
Y_{TF}= -8\pi\Pi +\frac{4\pi}{R^3}\int^r_0{R^3\left(D_R {\mu}-3{q}\frac{U}{RE}\right)R^\prime d r}.
\label{Y}
\end{equation}
Again,  we notice  that due to a different signature, the sign of $Y_{TF}$ in the above equation differs from the sign of the $Y_{TF}$ used in \cite{ch1} for the static case.

Thus the scalar $Y_{TF}$ may be expressed through the Weyl tensor and the anisotropy of pressure  or in terms of the anisotropy of pressure, the density inhomogeneity and  the dissipative variables.

Once the complexity factor for the structure of the fluid distribution has been established, it remains to elucidate what is the simplest pattern of evolution.
Based on purely intuitive thoughts we shall first identify  the homologous evolution as the simplest mode of evolution. 

In order to obtain a mathematical description of the homologous evolution, we shall proceed as follows 

First of all observe that  we can write (\ref{21a}) as
\begin{equation}
D_R\left(\frac{U}{R}\right)=\frac{4 \pi}{E} q+\frac{\sigma}{R},
\label{vel24}
\end{equation}
which after integration  becomes
\begin{equation}
U=\tilde a(t) R+R\int^r_0\left(\frac{4\pi}{E} q+\frac{\sigma}{R}\right)R^{\prime}dr,
\label{25n}
\end{equation}
where $\tilde a$ is an integration function, or,
\begin{equation}
U=\frac{U_{\Sigma}}{R_{\Sigma}}R-R\int^{r_{\Sigma}}_r\left(\frac{4\pi}{E} q+\frac{\sigma}{ R}\right)R^{\prime}dr.
\label{26n}
\end{equation}
If the integrand  in the above equations vanishes  we have from (\ref{25n}) or (\ref{26n}) 

\begin{equation}
 U= \tilde a(t) R.
 \label{ven6}
 \end{equation}
 
This relationship  is characteristic of the homologous evolution in Newtonian hydrodynamics \cite{22s,21s,20s}. In our case  this may occur if the fluid is shear--free and non dissipative, or if the two terms in the integral cancel each other.

In \cite{ch2}, the term  ``homologous evolution'' was used to characterize  relativistic systems satisfying, besides  (\ref{ven6}), the condition
\begin{equation}
\frac{R_I}{R_{II}}=\mbox{constant},
\label{vena}
\end{equation}
where $R_I$ and $R_{II}$ denote the areal radii of two concentric shells ($I,II$) described by $r=r_I={\rm constant}$, and $r=r_{II}={\rm constant}$, respectively. 

Now, it is very important to be aware of the fact that conditions (\ref{ven6}) and (\ref{vena}, in the general relativistic case, are different and more specifically,  that (\ref{ven6}) does not imply (\ref{vena}).

Indeed, (\ref{ven6})  implies that for the two shells of fluids $I,II$ we have 
\begin{equation}
\frac{U_I}{U_{II}}=\frac{A_{II} \dot R_I}{A_I \dot R_{II}}=\frac{R_I}{R_{II}},
\label{ven3}
\end{equation}
that implies (\ref{vena}) only if  $A=A(t)$, which by a simple coordinate transformation becomes $A={\rm constant}$. Thus in the non--relativistic regime, (\ref{vena}) always follows from the condition that  the radial velocity is proportional to the radial distance, whereas in the relativistic regime the condition (\ref{ven6}) implies (\ref{vena}), only if the fluid is geodesic. 

We shall define  quasi--homologous evolution as that restricted only by condition  (\ref{ven6}), implying 
\begin{equation}
\frac{4\pi}{R^\prime}B  q+\frac{\sigma}{ R}=0.
\label{ch1}
\end{equation}

Let us first consider the homologous evolution (both (\ref{ven6}) and (\ref{vena}) are satisfied).

From the equation  (\ref{vena}) it follows that $R$ is a separable function, i.e. we can write
\begin{equation}
 R=R_1(t) R_2(r).
 \label{hom2}
 \end{equation}

To summarize, the homologous condition   implies (\ref{hom2}), and 
\begin{equation}
\frac{4\pi}{R^\prime}B  q+\frac{\sigma}{ R}=0.
\label{ch1}
\end{equation}

Feeding back this last expression into (\ref{17a}), we obtain 
\begin{equation}
(\Theta-\sigma)^\prime=0,
\label{con5b}
\end{equation}
whereas, using (\ref{5c1}) and (\ref{5b1}) we get
\begin{equation}
\left(\Theta-\sigma\right)^\prime =\left(\frac{3}{A}\frac{\dot R}{R}\right)^\prime=0.
\label{con6b} 
\end{equation}
Then using (\ref{hom2}) it follows at once that 
\begin{equation}
A^\prime=0,
\label{con7}
\end{equation}
implying that  the fluid is geodesic, as it follows from (\ref{5c}). Also,  by reparametrizing the coordinate $r$, we may put, without loss of generality, $A=1$.

The inverse is in general untrue, e.g. the Lemaitre--Tolman-- Bondi case, unless we assume that  $(\Theta-\sigma)^\prime$ is of class $C^\omega$.

In the non--dissipative case, the homologous condition not only implies that the fluid is geodesic, but also that it is shear--free, as it follows at once from (\ref{ch1}).

An  important point  to mention here is that as it  has been  shown in \cite{26}, an initially  shear--free geodesic fluid remains shear--free during the evolution iff $Y_{TF}=0$. This implies that a system which  starts its evolution from the rest ($\sigma=0$),  will remain shear--free if  the fluid is geodesic (or equivalently, homologous) and $Y_{TF}=0$. This is an additional argument supporting our choice of $Y_{TF}$ as the complexity factor.

If we impose the homologous condition,  the equation (\ref{28}) in Appendix \ref{E} becomes
\begin{equation}
D_TU=-\frac{m}{R^2}-4\pi  P_r R
, \label{28bis}
\end{equation}
or in terms of $Y_{TF}$ 
\begin{equation}
\frac{3D_TU}{R}=-4\pi\left( \mu +3 P_r- 2\Pi\right)+Y_{TF},
 \label{29bis}
\end{equation}
where (\ref{EY}) has been used.
 
Next from the field equations we obtain

\begin{equation}
4\pi\left(\mu +3P_r- 2\Pi\right)=-\frac{2 \ddot R}{R}-\frac{ \ddot B}{B},
 \label{31bis}
\end{equation}
and from the definition of $U$
\begin{equation}
\frac{3D_TU}{R}=\frac{3 \ddot R}{R},
 \label{32bis}
\end{equation}
feeding back the two equations above into (\ref{29bis}), it follows that
\begin{equation}
\frac{ \ddot R}{R}-\frac{ \ddot B}{B}=Y_{TF}.
 \label{33bisa}
\end{equation}

Now, the vanishing complexity factor condition ($Y_{TF}=0$), produces after  the integration of  (\ref{33bisa}) 
\begin{equation}
B=R_1(t)\left(b_1(r)\int{\frac{dt}{R_1(t)^2}+b_2(r)}\right),
\label{int1}
\end{equation}
where $b_1(r)$ and $b_2(r)$ are two functions of integration, or 

\begin{equation}
B=R_1(t)R^\prime_2(r)\left(\tilde b_1(r)\int{\frac{dt}{R_1(t)^2}+\tilde b_2(r)}\right),
\label{int1bis}
\end{equation}
with $ b_1(r)=\tilde b_1(r)R_2^\prime$ and $ b_2(r)=\tilde b_2(r)R_2^\prime$.

Then introducing the variable 
\begin{equation}
Z=\tilde b_1(r)\int{\frac{dt}{R_1(t)^2}+\tilde b_2(r)},
\label{int13}
\end{equation}
we may write
\begin{equation}
B=ZR^\prime.
\label{int14}
\end{equation}
Let us now use the expressions above to analyze  first the non--dissipative case.

Thus, if we further assume  the fluid to be non--dissipative, recalling that in this case the homologous condition implies the vanishing of the shear, we obtain  because  of  (\ref{5b1}) 
\begin{equation}
\frac{ \ddot R}{R}-\frac{ \ddot B}{B}=0\quad \Rightarrow Y_{TF}=0.
 \label{33bis}
\end{equation}

In other words, in this particular case, the homologous condition already implies the vanishing complexity factor condition.

More so, since  the fluid is shear--free, we have because of (\ref{5b1}) and (\ref{int1})
\begin{equation}
b_1(r)=0 \Rightarrow B=b_2(r)R_1(t)=\tilde b_2(r)R_1(t)R_2^\prime.
 \label{34bis}
\end{equation}

Then reparametrising $r$ as $\tilde b_2(r)dr\Rightarrow dr$, we may put without loss of generality $B=R_1(t)R_2(r)^\prime$, or equivalently $Z=1$, implying that  all non--dissipative  configurations evolving homologously (and thereby satisfying $Y_{TF}=0$), belong to what are known as ``Euclidean stars'' \cite{euc}, characterized by the condition $Z=1 \Rightarrow B=R^\prime$. However among all possible solutions satisfying the ``Euclidean condition'' , only one evolves homologously and satisfies the condition $Y_{TF}=0$.

Indeed, from   the field equations (\ref{14}) and (\ref{15}) we may write
\begin{equation}
8\pi (P_r-P_{\bot})=\frac{\dot Z \dot R}{Z R}+\frac{1}{Z^2 R^2}\left(\frac{Z^\prime R}{ZR^\prime}+1-Z^2\right).
\label{necnd2}
\end{equation}

Since in this case we have $Z=1$ then $\Pi=P_r-P_\bot=0$ which implies because of the $Y_{TF}=0$ condition, that $\mu^\prime=0$. 

However it is known  that a shear--free, geodesic (non--dissipative) fluid with isotropic pressure is necessarily dust with homogeneous energy density and vanishing Weyl tensor (see \cite{gen,sc}). It goes without saying that this kind of system represents the simplest conceivable configuration (Friedman--Robertson--Walker).

Thus for the non--dissipative case, the homologous condition implies  $Y_{TF}=0$ and produces the simplest configuration. This configuration is the only one evolving homologously and satisfying  $Y_{TF}=0$. 

Of course, solutions satisfying $Y_{TF}=0$   but not evolving homologously do exist. They only require  $8\pi \Pi=\frac{4\pi}{R^3}\int^r_0{R^3 \mu^\prime dr}$. In such a case the solutions are shearing, and neither conformally flat nor geodesic. 

Based on all the precedent comments, it seems reasonable to  consider  the homologous condition  as a good candidate to describe the simplest  mode of evolution.

\subsection{The dissipative case}
In the dissipative case, we may obtain from (\ref{5b1}) and (\ref{33bis}), 

\begin{equation}
\dot \sigma =-Y_{TF}+\left(\frac{ \dot R}{R}\right)^2-\left(\frac{ \dot B}{B}\right)^2.
\label{37bis}
\end{equation}

Then, taking the $t$-derivative of (\ref{ch1}) and using (\ref{37bis}) we obtain
\begin{equation}
Y_{TF}\frac{R^\prime}{R}=4\pi Bq\left(\frac{\dot{q}}{q}+2\frac{\dot B}{B}+\frac{\dot R}{R}\right).
\label{38bis}
\end{equation}

If we assume $Y_{TF}=0$, then  we obtain
\begin{equation}
 q=\frac{f(r)}{B^2R},
\label{39bis}
\end{equation}
implying
\begin{equation}
 \dot q=-q(\Theta+\sigma),
\label{39bisb}
\end{equation}
where $f$ is an arbitrary integration function.
Solutions of this kind might be found by using the general methods presented in \cite{TM, TMb, ivanov0, ivanov1, ivanov}.

In the dissipative case we need to provide a transport equation to describe the evolution and  distribution of temperature.
Assuming a causal dissipative theory (e.g. the Israel-- Stewart theory \cite{19nt, 20nt, 21nt} ), the transport equation for the heat flux reads
\begin{equation}
\tau h^{\alpha \beta}V^\gamma q_{\beta;\gamma}+q^\alpha=-\kappa h^{\alpha \beta}\left(T_{,\beta}+Ta_\beta\right)-\frac{1}{2}\kappa T^2 \left(\frac{\tau V^\beta}{\kappa T^2}\right)_{;\beta} q^\alpha,
\label{tre}
\end{equation}
where $\kappa$ denotes the thermal conductivity, and $T$ and $\tau$ denote temperature and relaxation time, respectively. 

In the spherically symmetric case under consideration, the transport equation has only one independent component, which may be obtained from (\ref{tre}) by contracting with the unit spacelike vector $K^\alpha$, producing
\begin{equation}
\tau V^\alpha  q_{,\alpha}+q=-\kappa \left(K^\alpha T_{,\alpha}+T a\right)-\frac{1}{2}\kappa T^2\left(\frac{\tau V^\alpha}{\kappa T^2}\right)_{;\alpha} q.
\label{5}
\end{equation}

Sometimes, it is possible to simplify the equation above, assuming  the so-called truncated  transport equation when the last term in (\ref{tre}) may be neglected \cite{t8}, producing 
\begin{equation}
\tau V^\alpha  q_{,\alpha}+q=-\kappa \left(K^\alpha T_{,\alpha}+T a\right).
\label{5trun}
\end{equation}

Now, in a dissipative process,  it appears reasonable to consider  the stationary state, prevailing once the system has relaxed and transient phenomena have vanished,  as an  example of the simplest dissipative regime.  Thus, if we assume the stationary state (neglecting the relaxation time), then the transport equation (\ref{tre}) reads
\begin{equation}
q=-\frac{\kappa T^\prime}{B}.
\label{40bis}
\end{equation}

Combining the above equation with (\ref{39bis}) we obtain
\begin{equation}
T^\prime=-\frac{f(r)}{\kappa BR}.
\label{41bis}
\end{equation}

At this point, however, neither can we  provide solid arguments to support further the assumption about the vanishing of the relaxation time as an indicator of  minimum complexity of  the dissipative regime, nor can we prove that exact solutions of this kind exist.

So far we have assumed the homologous condition in order to describe the simplest mode of evolution, however as indicated in the lines above, the resulting models are perhaps too restrictive, and it could be wise to consider less stringent conditions. One possible example could be the quasi--homologous condition (\ref{ch1}). In \cite{rg5,rg6n,hcm} the reader may find a long list of exact solutions satisfying the vanishing complexity factor and the quasi--homologous condition for spherically symmetric and hyperbolically symmetric dissipative fluids.

Finally  we shall tackle the problem of defining complexity of vacuum solutions to Einstein equations. We shall restrict our discussion to the Bondi metric.

\section{Complexity of the  Bondi space--time}
Let us now consider the extension of our concept of complexity  to vacuum spacetimes. More specifically we shall consider the Bondi metric \cite{Boal}, which encompasses  a vast numbers of spacetimes, including the Minkowski spacetime, the static Weyl metrics, non--radiative non--static metrics and gravitationally radiating metrics.
Furthermore,  the Bondi approach has the virtue of providing a clear and precise criterion for the existence of gravitational
radiation. Namely, if the news function is zero over a time interval, then there
is no radiation over that interval. 

As we have seen, in the case of fluid distributions, the variable(s) measuring the complexity of the fluid (the complexity factor(s))  appear in the trace free part of the orthogonal splitting of the electric  Riemann tensor. In vacuum the Riemann tensor and the Weyl tensor are the same, so if we extrapolate to the vacuum case the same definition of complexity as for the fluid self--gravitating system, we shall need the scalar functions defining the electric part of the Weyl tensor for the Bondi metric. 

The general form of an axially and reflection symmetric asymptotically flat
metric given by Bondi \cite{Boal}  is  
\begin{eqnarray}
ds^2 & = & \left(\frac{V}{r} e^{2\beta} - U^2 r^2 e^{2\gamma}\right) du^2
+ 2 e^{2\beta} du dr  +  2 U r^2 e^{2\gamma} du d\theta \nonumber \\
&- &r^2 \left(e^{2 \gamma} d\theta^2 + e^{-2\gamma} \sin^2{\theta} 
d\phi^2\right),
\label{Bm}
\end{eqnarray}
where $V, \beta, U$ and $\gamma$ are functions of
$u, r$ and $\theta$.

We number the coordinates $x^{0,1,2,3} = u, r, \theta, \phi$ respectively.
$u$ is a timelike coordinate ($g_{uu}>0$ ) converging to the retarded time as $r\rightarrow \infty$. The hypersurfaces  $u=constant$ define  null surfaces (their normal vectors  are  null vectors), which at null infinity ($r\rightarrow \infty$) coincides with the Minkowski null light cone
open to the future. $r$ is a null coordinate ($g_{rr}=0$) and $\theta$ and
$\phi$ are two angle coordinates (see \cite{Boal} for details).

Regularity conditions in the neighborhood of the polar axis
($\sin{\theta}=0$), imply that
as $\sin{\theta} \rightarrow 0$
\begin{equation}
V, \beta, U/\sin{\theta}, \gamma/\sin^2{\theta},
\label{regularity}
\end{equation}
each equals a function of $\cos{\theta}$ regular on the polar axis.

The four metric functions are assumed to be expanded in series of $1/r$,
then using the field equations, Bondi gets

\begin{equation}
\gamma = c r^{-1} + \left(C - \frac{1}{6} c^3\right) r^{-3}
+ ...,
\label{ga}
\end{equation}
\begin{equation}
U = - \left(c_\theta + 2 c \cot{\theta}\right) r^{-2} + \left[2
N+3cc_{\theta}+4c^2 \cot{\theta}\right]r^{-3}...,
\label{U}
\end{equation}

\begin{eqnarray}
V  = r - 2 M
 -  \left( N_\theta + N \cot{\theta} -
c_{\theta}^{2} - 4 c c_{\theta} \cot{\theta} -\right. \\ \left.
\frac{1}{2} c^2 (1 + 8 \cot^2{\theta})\right) r^{-1}  ...,
\label{V}
\end{eqnarray}
\begin{equation}
\beta = - \frac{1}{4} c^2 r^{-2} + ...,
\label{be}
\end{equation}
 where $c$, $C$, $N$ and $M$ are functions of $u$ and $\theta$ satisfying the constraint

\begin{equation}
4C_u = 2 c^2 c_u + 2 c M + N \cot{\theta} - N_\theta,
\label{C}
\end{equation}
and letters as
subscripts denote derivatives.
The three functions $c, M$ and $N$ are further
related by the supplementary conditions
\begin{equation}
M_u = - c_u^2 + \frac{1}{2}
\left(c_{\theta\theta} + 3 c_{\theta} \cot{\theta} - 2 c\right)_u,
\label{Mass}
\end{equation}
\begin{equation}
- 3 N_u = M_\theta + 3 c c_{u\theta} + 4 c c_u \cot{\theta} + c_u c_\theta.
\label{N}
\end{equation}

In the static case $M$ equals the mass of the system and is called by Bondi the ``mass aspect'', whereas $N$ and $C$
are closely related to the dipole and quadrupole moments respectively.

Next, Bondi defines the mass $m(u)$ of the system as
\begin{equation}
m(u) = \frac{1}{2} \int_0^\pi{M \sin{\theta} d\theta},
\label{m}
\end{equation}
which by virtue of (\ref{Mass}) and (\ref{regularity}) yields
\begin{equation}
m_u = - \frac{1}{2} \int_0^\pi{c_u^2 \sin{\theta} d\theta}.
\label{muI}
\end{equation}

Arriving at this point let us summarize  the main conclusions emerging from  Bondi's approach.
\begin{enumerate}
\item If $\gamma, M$ and $N$ are known for some $u=a$(constant) and
$c_u$ (the news function) is known for all $u$ in the interval
$a \leq u \leq b$,
then the system is fully determined in that interval. In other words,
whatever happens at the source, leading to changes in the field,
it can only do so by affecting $c_u$ and vice versa. In the
light of this comment the relationship between news function
and the occurrence of radiation becomes clear.
\item As it follows from (\ref{muI}), the mass of a system is constant
if and only if there is no news.
\end{enumerate}

Now, for an observer at rest in the frame of (\ref{Bm}), the four-velocity
vector has components
\begin{equation}
V^{\alpha} = \left(\frac{1}{A}, 0, 0, 0\right),
\label{fvct}
\end{equation}
with
\begin{equation}
A \equiv \left(\frac{V}{r} e^{2\beta} - U^2 r^2 e^{2\gamma}\right)^{1/2}.
\label{A}
\end{equation}

Next, let us  introduce the unit, spacelike vectors ${\bf K}$, ${\bf L}$, ${\bf S}$, with components
\begin{equation}
K^{\alpha} = \left(\frac{1}{A}, -e^{-2\beta}A, 0, 0\right)\quad L^{\alpha} = \left(0, Ure^{\gamma}e^{-2\beta}, -\frac{e^{-\gamma}}{r}, 0\right)
\label{K}
\end{equation}
\begin{equation}
S^{\alpha} = \left(0, 0, 0, -\frac{e^{\gamma}}{r\sin \theta} \right),
\label{K}
\end{equation}

For the observer defined by  (\ref{fvct}) the vorticity vector may be
written as (see \cite{HSC} for details)
\begin{equation}
\omega^\alpha = \left(0, 0, 0, \omega^{\phi}\right).
\label{oma}
\end{equation}
The explicit expressions  for $ \omega^{\phi}$ and its absolute value $\Omega  \equiv  \left(- \omega_\alpha \omega^\alpha\right)^{1/2}$ are given in the Appendix \ref{H}.

The electric and magnetic parts of Weyl tensor, $E_{\alpha \beta}$ and
$H_{\alpha\beta}$, respectively, are formed from the Weyl tensor $C_{\alpha
\beta \gamma \delta}$ and its dual
$\tilde C_{\alpha \beta \gamma \delta}$ by contraction with the four
velocity vector given by (\ref{fvct})
\begin{equation}
E_{\alpha \beta}=C_{\alpha \gamma \beta \delta}V^{\gamma}V^{\delta},
\label{electric}
\end{equation}
\begin{eqnarray}
H_{\alpha \beta}&=&\tilde C_{\alpha \gamma \beta \delta}V^{\gamma}V^{\delta}=
\frac{1}{2}\epsilon_{\alpha \gamma \epsilon \delta} C^{\epsilon
\delta}_{\quad \beta \rho} V^{\gamma}
V^{\rho},\nonumber \\
&& \epsilon_{\alpha \beta \gamma \delta} \equiv \sqrt{-g}
\;\;\eta_{\alpha \beta \gamma \delta},
\label{magnetic}
\end{eqnarray}
where
$\eta_{\alpha\beta\gamma\delta}$ is the permutation symbol.

The electric part of the Weyl tensor has only three independent non-vanishing components, whereas only two components define the magnetic part. Thus  we may  write

\begin{eqnarray}
E_{\alpha \beta}&=&\mathcal{E}_1\left(K_\alpha L_\beta+L_\alpha K_\beta\right)
+\mathcal{E}_2\left(K_\alpha K_\beta+\frac{1}{3}h_{\alpha \beta}\right)\nonumber \\&+&\mathcal{E}_3\left(L_\alpha L_\beta+\frac{1}{3}h_{\alpha \beta}\right), \label{13}
\end{eqnarray}

and
\begin{equation}
H_{\alpha\beta}=H_1(S_\alpha K_\beta+S_\beta
K_\alpha)+H_2(S_\alpha L_\beta+S_\beta L_\alpha)\label{H'}.
\end{equation}

with $h_{\mu \nu}=g_{\mu\nu}-V_\nu V_\mu$, and

\begin{equation}
\mathcal{E}_1=L^\alpha K^\beta E_{\alpha \beta},
\label{ew3}
\end{equation}
\begin{equation}
\mathcal{E}_2=(2K^\alpha K^\beta+ L^\alpha L^\beta)E_{\alpha \beta},
\label{ew4}
\end{equation}
\begin{equation}
\mathcal{E}_3=(2L^\alpha L^\beta+ K^\alpha K^\beta)E_{\alpha \beta},
\label{ew4}
\end{equation}
these three scalars represent  the complexity  factors of  our solutions.

For the magnetic part we have
\begin{equation}
H_2=S^\alpha L^\beta H_{\alpha \beta},
\label{ew5}
\end{equation}
 \begin{equation}
H_1=S^\alpha K^\beta H_{\alpha \beta}.
\label{ew5}
\end{equation}
 
Explicit expressions for these scalars are given in the Appendixes \ref{F} and \ref{G}.

In \cite{HSC} it was obtained that if we put $H^{\alpha}_{\beta}=0$ then
the field is non--radiative and 
 up to order $1/r^3$ in $\gamma$, the metric is static, and the
mass, the ``dipole'' ($N$) and the ``quadrupole'' ($C$) moments correspond
to a static situation. However, the
time dependence might enter through coefficients of higher order in
$\gamma$, giving rise to what Bondi calls ``non--natural--non--radiative
moving system'' (NNNRS). In this latter case,
the system keeps the first three  moments independent of time, but allows
for time dependence of higher moments.  This class of solutions is characterized by $M_{\theta}=0$. 

A second family of time dependent non--radiative solutions exists for which $M_{\theta}\neq 0$. These are called natural non--radiative
moving system'' (NNRS), and their magnetic Weyl tensor is non--vanishing.

Let us now  discuss the hierarchy of different spacetimes belonging to the Bondi family, according to their complexity.

The simplest spacetime corresponds to the vanishing of the three complexity factors, and this is just Minkowski.

Indeed, as it was shown in \cite{HSC},  if we assume
$E^{\alpha}_{\beta}=0$ and use regularity conditions, we find that  the spacetime must be Minkowski,
giving further support to the
conjecture that there are no purely magnetic vacuum space--times 
\cite{Bonnor}.

At the other end (maximal complexity) we have a gravitationally radiating system which requires all three complexity factors to be different from zero.

Indeed, if we  assume that $\mathcal{E}_1=0$, then it follows at once from (\ref{e1}) that $c_u=0$ (otherwise $c_u$ would be a non--regular function of $\theta$ on the symmetry axis). Thus $\mathcal{E}_1=0$ implies
 that the system is non--radiative.
 
 If instead we assume that $\mathcal{E}_2=0$, then from the first order in (\ref{e2a}) we obtain that $c_{uu}=0$, this implies that either $c_u=0$ or $c_u\sim u$. Bondi refers to this latter case as ``mass loss without radiative Riemann tensor'' and dismisses it as being of little physical significance. As a matter of fact, in this latter case the system would be radiating ``forever'', which  according to (\ref{muI}) requires an unbounded source, incompatible with an asymptotically flat spacetime. Thus in this case too, we have $c_u=0$, and the system is non--radiative.
 
 Finally, if we assume $\mathcal{E}_3=0$ it follows at once from the first order in (\ref{e3}), that $c_{uu}=0$, leading to $c_{u}=0$, according to the argument above.

Thus, a radiative system requires all  three complexity factors to be nonvanishing, implying  a maximal complexity.

In the middle of the two extreme cases   we have, on the one hand  the spherically symmetric spacetime (Schwarzschild), characterized by a single complexity factor (the same applies for any static metric), $\mathcal{E}_1=\mathcal{E}_3=0$, and $\mathcal{E}_2=\frac{3M}{r^3}$. On the other hand, we have the non--static non--radiative case.

Let us now analyze in detail this latter case. This group of solutions encompasses two subclasses, which using Bondi notation are:
\begin{enumerate}
\item Natural--non--radiative systems (NNRS) characterized by $M_\theta\neq0$.
\item Non--natural--non--radiative systems (NNNRS) characterized by $M_\theta=0$.
\end{enumerate}

Let us first consider the NNNRS subcase. Using (\ref{e1}) we obtain  ${\cal E}_1=0$, (up to order $1/r^3$),  while the first non--vanishing terms in ${\cal E}_2$ and ${\cal E}_3$ are respectively,
$3M$ and $0$,
where (\ref{C}), (\ref{Mass}), (\ref{N}), (\ref{e2a}) and (\ref{e3}) have been used. 

Thus,  the NNNRS are characterized by only one non--vanishing complexity factor (${\cal E}_2$).  Furthermore, as it follows from (\ref{OM}) the vorticity of the congruence of observers at rest with respect to the frame of (\ref{Bm}) vanishes, and the field is purely electric. However as mentioned before we cannot conclude that the field is static, since the $u$ dependence might appear through coefficients of higher order in $\gamma$.

Let us now consider the  ``natural--non--radiative  system'' (NNRS). In this subcase, using (\ref{e1}) we obtain  ${\cal E}_1=0$, (up to order $1/r^3$) as for the NNNRS subcase, while the first non--vanishing term in ${\cal E}_2$ and ${\cal E}_3$  (up to order $1/r^3$)  are  respectively $3M+\frac{M_{\theta \theta}}{4}-\frac{M_{ \theta}\cot \theta}{4}$ and  $\frac{M_{\theta \theta}}{2}-\frac{M_{ \theta}\cot \theta}{2}$. 

Also, up to the same order, it follows from (\ref{h1}) and (\ref{h2}) that $H_1=0$ for both subcases, while the corresponding term in $H_2$ is (for NNRS)
\begin{equation}
-\frac{1}{4}(M_{\theta \theta}-M_\theta \cot\theta),
\label{h2r3}
\end{equation}
which of course vanishes for the NNNRS subcase.

It should be observed that if we assume  ${\cal E}_3=0$ or $H_2=0$  then it follows at once from the above that 

\begin{equation}
M_{\theta \theta}-M_\theta \cot\theta=0 \Rightarrow M=a\cos\theta, \quad a=constant.
\label{h2r3f}
\end{equation}

But this implies because of   (\ref{m}) that the Bondi mass function of the system vanishes. Therefore, the only physically meaningful NNRS requires ${\cal E}_3\neq0$, $\Omega\neq0$  and $H_2\neq0$  implying that the complexity is characterized by two complexity factors (${\cal E}_2$, ${\cal E}_3$).

Thus  a hierarchy of  spacetimes according to their complexity has been established. This allows us  to discriminate between  two classes of spacetimes that depend on time but are not radiative (vanishing of the news function).  These two classes were called by Bondi \cite{Boal} natural and non--natural non--radiative moving systems, and are characterized by different forms  of the mass aspect. They exhibit different degree of complexity. 

Unfortunately, though, up to the leading order of the complexity factors analyzed here, it is  impossible to discriminate between different radiative systems according to their complexity. Higher order terms would be necessary for that purpose, although it is not clear at this point if it is possible to establish such a hierarchy of radiative systems after all.

The simplest system (Minkowski) is characterized by the vanishing of all the complexity factors. Next, the static case (including Schwarzchild) is described by a single complexity factor. 

The time dependent non--radiative solutions split in two subgroups depending on the form of the mass aspect $M$. If $M_\theta=0$ which corresponds to the NNNRS the complexity is similar to the static case.  Also, in this case, as in the static situation, the vorticity vanishes and the field is purely electric. This result could  suggest that in fact  NNNRS are just static, and no time dependence appears  in the coefficients of higher order in
$\gamma$. On the contrary for the NNRS there are two complexity factors, the vorticity is non--vanishing and  the field is not purely electric. 

All these results are summarized  in Tables I and II.
Thus, NNNRS and NNRS are clearly differentiated through their degree of complexity, as measured by the complexity factors considered here. 

The fact that radiative systems necessarily decay into NNRS, NNNRS or static systems, since  the Bondi mass function must be finite, suggests that higher degrees of complexity might be  associated with stronger stability. Of course a proof of this conjecture requires a much more detailed analysis.

It is also worth mentioning  the conspicuous link between vorticity and complexity factors. Indeed vorticity appears only in NNRS and radiative systems, which are the most complex systems, while it is absent in the simplest systems (Minkowski, static, NNNRS). In the radiative case there are contributions at order $\mathcal{O}(r^{-1})$  related to the news function, and at  order $\mathcal{O}(r^{-2})$, while for  the NNRS there are only contributions at order $\mathcal{O}(r^{-2})$, these  describe the effect of the tail of the wave.
\begin{widetext}
\begin{table}[htp]
\caption{Complexity factors for different spacetimes of the Bondi metric}
\begin{center}
COMPLEXITY HIERARCHY
 
\begin{tabular}{ccccccccccccccccc}
\hline\hline
$complex. fac. \diagdown    spacetimes$ &$\quad$&
Minkowski &$\quad$ &
Static &$\quad$ &
NNNRS &$\quad$ &
NNRS &$\quad$ &
Radiative &$\quad$ &
 \vspace{0.3cm} \\  \hline
${\cal E}_1$&$\quad$ &0  &$\quad$ &  0     &$\quad$ & 0   &$\quad$ &0 &$\quad$ & ${\cal E}^{(n)}_1\neq 0$, $n\geq 1$\\ \hline
${\cal E}_2$ &$\quad$ & 0  & &     ${\cal E}^{(3)}_2=3M$& & ${\cal E}^{(3)}_2=3M$&  & ${\cal E}^{(3)}_2=3M+\frac{M_{\theta \theta}}{4}-\frac{M_{ \theta}\cot \theta}{4}$&$\quad$ & ${\cal E}^{(n)}_2\neq 0$, $n\geq 1$ \\ \hline
${\cal E}_3$&$\quad$ & 0  &$\quad$ &   0       &$\quad$ & 0  &$\quad$ & ${\cal E}^{(3)}_3=\frac{1}{2}(M_{\theta \theta}-M_{ \theta}\cot \theta)$&$\quad$ & ${\cal E}^{(n)}_3\neq 0$, $n\geq 1$ \\ \hline
\end{tabular}
\end{center}
\label{data}
\end{table}
\end{widetext}
where ${\cal E}^{(n)}_{1,2,3}$ are de coefficients of order $\mathcal{O}(r^{-n})$.

\begin{widetext}
\begin{table}[htp]
\caption{The magnetic parts of the Weyl tensor and the vorticity for different spacetimes of the Bondi metric}
\begin{center}
Magnetic parts and vorticity
 
\begin{tabular}{ccccccccccccccccc}
\hline\hline
$Mag. Weyl; \Omega. \diagdown    spacetimes$ &$\quad$&
Minkowski &$\quad$ &
Static &$\quad$ &
NNNRS &$\quad$ &
NNRS &$\quad$ &
Radiative &$\quad$ &
 \vspace{0.3cm} \\  \hline
$H_1$&$\quad$ &0  &$\quad$ &  0     &$\quad$ & 0   &$\quad$ &0 &$\quad$ & $H^{(n)}_1\neq 0$, $n\geq 1$\\ \hline
$H_2$ &$\quad$ & 0  &$\quad$ &   $0$      &$\quad$ & $0$   &$\quad$ & $H^{(3)}_2=-\frac{1}{4}(M_{\theta \theta}-M_{ \theta}\cot \theta)$ &$\quad$ & $ H^{(n)}_2\neq 0$, $n\geq 1$  \\ \hline
$\Omega$&$\quad$ & 0  &$\quad$ &   0       &$\quad$ & 0  &$\quad$ & $\Omega^{(2)}=M_{ \theta}$&$\quad$ & $\Omega^{(n)}\neq 0$, $n\geq 1$ \\ \hline
\end{tabular}
\end{center}
\label{data}
\end{table}
 \end{widetext}

\section{Conclusions}
We have discussed about the concept  of complexity of self--gravitating  relativistic fluid distributions. For the static case we were concerned exclusively with the notion of complexity of the structure of the fluid. However in the dynamic case we have also tackled the question about the complexity of the pattern of evolution. These two  questions,  although related, refer to different aspects of  the definition of complexity. However, it is remarkable that in the non--dissipative case the homologous condition implies the vanishing complexity factor.

As a measure of complexity of the structure of the fluid (the complexity factor) we have chosen the scalar function(s)  $Y_{TF}$ defining the trace--free part of the electric Riemann tensor. 

Next, we discussed about the complexity of the pattern of evolution. Two possibilities appear as the more obvious candidates: the homologous condition and the quasi--homologous condition. The latter, being less restricted than the former allows to consider  larger number of models.

All this having been said, we are well aware of the fact that the above mentioned candidate for measuring the complexity  of a self-gravitating system is by no means unique and many different alternatives may be proposed.  In the same line of arguments, the simplest patterns of evolution assumed so far are the homologous and  the quasi-homologous regimes. But, it is not clear whether or not  other patterns of evolution could  also fit the role of the simplest pattern of evolution. 

Additionally, we believe that alternative   definitions of  complexity for vacuum space--time is worth considering. Finally, new exact solutions to the field equations in the context of the Einstein theory or any alternative one, would serve as a test-bed for the definition of complexity.

These remarks suggest  a list of questions and open issues which we believe that deserve to be considered further.

\begin{itemize}
\item Are there alternative definitions of complexity, different from the one  proposed in \cite{ch1}?
\item Are there other ways to  extend the definition of complexity for vacuum space--time?
\item Besides the homologous and the quasi-homologous regime, could we define another pattern  of evolution that could qualify as the simplest one?
\item Do physically meaningful dissipative models satisfying  (\ref{39bis}) exist?
\item If the answer to the above question is positive then, is there a unique solution or  are there a large number of them?
\item What is the physical meaning of such solution(s)?
\item Is it physically reasonable to neglect transient effects when considering the simplest dissipative system, and assume that the relaxation time vanishes?
\item To summarize the four  points above: is there a specific dissipative regime that could be considered as the simplest one?
\item Can we relate the complexity factor(s) in the non-spherically symmetric case to the active gravitational mass, as in the spherically symmetric case?
\item Can we single out a specific family of exact axially symmetric static solutions satisfying the vanishing complexity factor(s) condition?
\item Can any of the above solutions be matched smoothly to any vacuum Weyl solution?
 \item The definition of complexity proposed in \cite{ch1} is not directly related to entropy or disequilibrium, although it is possible that such a link might exist after all. If so, how could such relationship be brought out?
 \item What is the simplest mode of evolution of axially--symmetric fluids? How is such a mode related to the emission (or not) of gravitational radiation?
 \item Could it be possible to provide a definition of the arrow of time in terms of the complexity factor?
\item How is the complexity factor related to physical relevant properties of the source, in terms of stability or  maximal degree of compactness?
\item  How does the complexity factor evolve? Do physically meaningful systems prefer vanishing complexity factors?
\item As we have seen the FRW model satisfies the vanishing complexity factor and evolves homologously.  Should any other,  physically sound, cosmological model have a vanishing complexity factor?  Should it evolve in the  homologous or quasi-homologous regime?
\item  The complexity factor for a charged fluid is known, but what is the complexity factor for a different type of field (e.g.,  scalar field?).
\item  How should we define the complexity factor  in the context of  other alternative theories of gravity that have not been considered so far?
\item How can we find new solutions satisfying the vanishing complexity factor? Could we use the general methods described in \cite{TM,TMb,ivanov,ivanov0,ivanov1} to obtain such solutions?
\item What relevant physical features share solutions satisfying the vanishing complexity factor?
\item Is there  a link between the concept of complexity and some kind of symmetry (e.g. motions, conformal motions, affine collineations, curvature collineations, matter collineations, etc.)?
\item We have  extended  the concept of complexity, adopted for fluids, to the vacuum case for the Bondi metric, the three complexity factors corresponding to the three scalars defining the electric part of the Weyl tensor. Could it be possible to further refine the scheme proposed here so as to discriminate between different radiative systems according to their complexity? Or, in other words, among  radiative systems is there a simplest one ? 
\item Also in the vacuum case, could it be  possible to discriminate between different static spacetimes of the Weyl family?
\end{itemize}

\section*{Acknowledgments}
This work was partially supported by Grant PID2021-122938NB-I00 funded by MCIN/AEI/ 10.13039/501100011033 and by  ERDF A way of making Europe. I also wish to thank  Universitat de les  Illes Balears   for financial support and hospitality. 

\section*{Appendix}
\addcontentsline{toc}{section}{Appendix}

\section{The Einstein and the conservation equations for the axially symmetric static case}
\label{A}
For the line element (\ref{1b}) and the energy momentum (\ref{6bisax}), the Einstein equations read:
\begin{widetext}
\begin{eqnarray}
8\pi\mu=-\frac{1}{B^2}\left\{\frac{B^{\prime \prime}}{B}+\frac{D^{\prime \prime}}{D}+\frac{1}{r}\left(\frac{B^\prime}{B} +\frac{D^\prime}{D}\right)-\left(\frac{B^\prime}{B}\right)^2  +\frac{1}{r^2}\left[\frac{B_{\theta \theta}}{B}+\frac{D_{\theta \theta}}{D}-\left(\frac{B_\theta}{B}\right)^2\right] \right\},
\label{24}
\end{eqnarray}

\begin{eqnarray}
8\pi P_{xx}=\frac{1}{B^2}\left[\frac{A^\prime B^\prime}{AB}+ \frac{A^\prime D^\prime}{AD}+\frac{B^\prime D^\prime}{BD}+\frac{1}{r}\left(\frac{A^\prime}{A}+\frac{D^\prime}{D}\right)+ \frac{1}{r^2}\left(\frac{A_{\theta \theta}}{A}+\frac{D_{\theta \theta}}{D}-\frac{A_\theta B_\theta}{AB}+\frac{A_\theta D_\theta}{AD}-\frac{B_\theta D_\theta}{BD}\right)\right],
\label{25}
\end{eqnarray}
\begin{eqnarray}
8\pi P_{yy}=\frac{1}{B^2}\left[\frac{A^{\prime \prime}}{A}+ \frac{D^{\prime \prime}}{D}-\frac{A^\prime B^\prime}{AB} +\frac{A^\prime D^\prime}{AD}-\frac{B^\prime D^\prime}{BD} +\frac{1}{r^2}\left(\frac{A_\theta B_\theta}{AB}+\frac{A_\theta D_\theta}{AD}+\frac{B_\theta D_\theta}{BD}\right)\right],
\label{27}
\end{eqnarray}

\begin{eqnarray}
8\pi P_{zz}=\frac{1}{B^2}\left\{\frac{A^{\prime \prime}}{A}+ \frac{B^{\prime \prime}}{B}-\left(\frac{B^\prime}{B}\right)^2+\frac{1}{r}\left(\frac{A^\prime}{A} +\frac{B^\prime}{B}\right) +\frac{1}{r^2}\left[\frac{A_{\theta \theta}}{A}+\frac{B_{\theta \theta}}{B}-\left(\frac{B_\theta}{B}\right)^2\right]\right\},
\label{28}
\end{eqnarray}

\begin{eqnarray}
8\pi P_{xy}=\frac{1}{B^2}\left\{  \frac{1}{r}\left[-\frac{A^{\prime}_\theta}{A} -\frac{D^{\prime}_\theta}{D} +\frac{B_\theta}{B}\left(\frac{A^\prime}{A}+\frac{D^\prime}{D}\right)+\frac{B^\prime}{B}\frac{A_\theta}{A}+\frac{B^\prime}{B}\frac{D_\theta}{D}\right]+\frac{1}{r^2} \left(\frac{A_\theta}{A}+\frac{D_\theta}{D}\right)\right\}.\label{26}
\end{eqnarray}
 \end{widetext}

The nonvanishing components of the conservation equations $T^{\alpha  \beta}_{;\beta}=0$ yield: the trivial equation
\begin{equation}
\dot \mu=0,
\label{21}
\end{equation}
where the overdot denotes derivative with respect to $t$,
and the two hydrostatic equilibrium equations 
\begin{widetext}
\begin{eqnarray}
\left[P+\frac{1}{3}(2\Pi_2-\Pi_3)\right]^\prime +\frac{A^{\prime}}{A}\left[\mu+P+\frac{1}{3}(2\Pi_2-\Pi_3)\right]+\frac{B^{\prime}}{B}(\Pi_2-\Pi_3)+\frac{D^{\prime}}{D}\Pi_2 \nonumber  \\+ \frac{1}{r}\left[\left(\frac{A_\theta}{A}+2\frac{B_\theta}{B}+\frac{D_\theta}{D}\right)\Pi_{KL}+\Pi_{Kl\theta}+\Pi_2-\Pi_3\right]=0,
\label{22}
\end{eqnarray}
\end{widetext}
\begin{widetext}
\begin{eqnarray}
\left[P+\frac{1}{3}(2 \Pi_3-\Pi_2)\right]_\theta+\frac{A_{\theta}}{A}\left[\mu +P+\frac{1}{3}(2 \Pi_3-\Pi_2)\right]+\frac{B_{\theta}}{B}(\Pi_3-\Pi_2)\nonumber \\+\frac{D_{\theta}}{D}\Pi_3+r\left[\left(\frac{A^{\prime}}{A}+2\frac{B^{\prime}}{B}+\frac{D^{\prime}}{D}\right)\Pi_{KL}+\Pi^{\prime}_{KL}\right]+2\Pi_{KL}=0.
\label{23}
\end{eqnarray}
\end{widetext}
\section{Expression for the components of the electric Weyl tensor}
\label{B}
There are four  nonvanishing components as calculated from (\ref{8}), however they are not independent since they satisfy the relationship:
\begin{equation}
E_{11}+\frac{1}{r^2}E_{22}+\frac{B^2}{D^2}E_{33}=0,
\end{equation}
implying that the Weyl tensor may be expressed through three independent scalar functions $\mathcal{E}_1, \mathcal{E}_2, \mathcal{E}_3$.

These four components are
\begin{widetext}
\begin{eqnarray}
E_{11} &=& \frac{1}{6}\left[\frac{2A^{\prime \prime}}{A}-\frac{B^{\prime \prime}}{B}-\frac{D^{\prime \prime}}{D}-\frac{3A^{\prime} B^{\prime}}{AB}-\frac{A^{\prime} D^{\prime}}{AD}+\left(\frac{B^{\prime}}{B}\right)^2+\frac{3B^{\prime} D^{\prime}}{BD}+\frac{1}{r}\left(2\frac{D^\prime}{D}-\frac{B^\prime}{B}-\frac{A^\prime}{A}\right)\right]\nonumber \\
&+&\frac{1}{6r^2}\left[-\frac{A_{\theta \theta}}{A} -\frac{B_{\theta \theta}}{B} +\frac{2 D_{\theta \theta}}{D} +\frac{3A_\theta B_\theta}{AB} -\frac{A_\theta D_\theta}{AD}+\left(\frac{B_\theta}{B}\right)^2-\frac{3B_\theta D_\theta}{BD}\right],\label{9}
\end{eqnarray}

\begin{eqnarray}
E_{22} &=& -\frac{r^2}{6}\left[\frac{A^{\prime \prime}}{A}+\frac{B^{\prime \prime}}{B}-\frac{2D^{\prime \prime}}{D}-\frac{3A^{\prime} B^{\prime}}{AB}+\frac{A^{\prime} D^{\prime}}{AD}-\left(\frac{B^{\prime}}{B}\right)^2+\frac{3B^{\prime} D^{\prime}}{BD}+\frac{1}{r}\left(\frac{D^\prime}{D}+\frac{B^\prime}{B}-\frac{2A^\prime}{A}\right)\right]\nonumber \\
&-&\frac{1}{6}\left[-\frac{2A_{\theta \theta}}{A} +\frac{B_{\theta \theta}}{B} +\frac{ D_{\theta \theta}}{D} +\frac{3A_\theta B_\theta}{AB} +\frac{A_\theta D_\theta}{AD}-\left(\frac{B_\theta}{B}\right)^2-\frac{3B_\theta D_\theta}{BD}\right],\label{10}
\end{eqnarray}

\begin{eqnarray}
E_{33} &=& -\frac{D^2}{6B^2}\left[\frac{A^{\prime \prime}}{A}-\frac{2B^{\prime \prime}}{B}+\frac{D^{\prime \prime}}{D}-\frac{2A^{\prime} D^{\prime}}{AD}+2\left(\frac{B^{\prime}}{B}\right)^2+\frac{1}{r}\left(\frac{D^\prime}{D}-\frac{2B^\prime}{B}+\frac{A^\prime}{A}\right)\right]\nonumber \\
&-&\frac{D^2}{6B^2r^2}\left[\frac{A_{\theta \theta}}{A} -\frac{2B_{\theta \theta}}{B} +\frac{ D_{\theta \theta}}{D}  -\frac{2A_\theta D_\theta}{AD}+2\left(\frac{B_\theta}{B}\right)^2\right],\label{11}
\end{eqnarray}

\begin{eqnarray}
E_{12}  = \frac{1}{2} \left[\frac{A^{\prime}_\theta}{A} -\frac{D^{\prime}_\theta}{D}+\frac{B_\theta}{B}\frac{D^{\prime}}{D}-\frac{A^\prime B_{\theta}}{AB}-\frac{B^\prime A_{\theta}}{AB}+\frac{D_\theta}{D}\frac{B^{\prime}}{B}-\frac{1}{r}\left(\frac{A_\theta}{A}-
\frac{D_\theta}{D}\right)\right] \label{12}.
\end{eqnarray}
\end{widetext}

For the three scalars $\mathcal{E}_1$, $\mathcal{E}_2$, $\mathcal{E}_3$  we obtain
\begin{widetext}
\begin{eqnarray}
\mathcal{E}_1= \frac{1}{2B^2} \left[\frac{1}{r}\left(\frac{A^{\prime}_\theta}{A} -\frac{D^{\prime}_\theta}{D}-
\frac{B_\theta}{B}\frac{A^{\prime}}{A}+\frac{D^{\prime}}{D}\frac{B_{\theta}}{B}-\frac{B^\prime}{B}\frac{A_\theta}{A}+
\frac{D_\theta}{D}\frac{B^\prime}{B}\right)+\frac{1}{r^2}\left(\frac{D_{\theta}}{D}-\frac{A_\theta}{A}\right)\right],\label{15}
\end{eqnarray}

\begin{eqnarray}
\mathcal{E}_2 & = &-\frac{1}{2B^2}\left[-\frac{A^{\prime \prime}}{A}+\frac{B^{\prime \prime}}{B}+
\frac{A^\prime B^\prime}{AB}+\frac{A^\prime D^\prime}{AD}-\left(\frac{B^\prime}{B}\right)^2-\frac{B^\prime D^\prime}{BD}+\frac{1}{r}\left(\frac{B^\prime}{B}
-\frac{D^\prime}{D}\right)\right]\nonumber \\ &-&\frac{1}{2B^2r^2}\left[\frac{B_{\theta \theta}}{B} -
\frac{D_{\theta \theta}}{D} -\frac{A_{\theta}B_{\theta}}{AB} +\frac{A_\theta D_\theta}{AD} -
\left(\frac{B_\theta}{B}\right)^2+\frac{B_\theta D_\theta}{BD}\right],\label{16}
\end{eqnarray}

\begin{eqnarray}
\mathcal{E}_3 & = &-\frac{1}{2B^2}\left[\frac{B^{\prime \prime}}{B}-\frac{D^{\prime \prime}}{D}-
\frac{A^\prime B^\prime}{AB}+\frac{A^\prime D^\prime}{AD}-\left(\frac{B^\prime}{B}\right)^2+\frac{B^\prime D^\prime}{BD}+\frac{1}{r}\left(\frac{B^\prime}{B}
-\frac{A^\prime}{A}\right)\right]\nonumber \\ &-&\frac{1}{2B^2r^2}\left[\frac{B_{\theta \theta}}{B} -\frac{A_{\theta \theta}}{A}
 +\frac{A_{\theta}B_{\theta}}{AB} +\frac{A_\theta D_\theta}{AD} -
\left(\frac{B_\theta}{B}\right)^2-\frac{B_\theta D_\theta}{BD}\right].\label{17}
\end{eqnarray}
\end{widetext} 

or using Einstein equations 
\begin{widetext}
\begin{eqnarray}
\mathcal{E}_1 =\frac{E_{12}}{B^2r}=4\pi \Pi_{KL}+\frac{1}{B^2 r}\left[\frac{A^{\prime}_\theta}{A}-
\frac{A^{\prime} B_\theta}{AB}-\frac{A_\theta}{A}\left (\frac{B^{\prime}}{B}+\frac{1}{r}\right)\right],
\label{18}
\end{eqnarray}

\begin{eqnarray}
\mathcal{E}_2 &=&-\frac{2E_{33}}{D^2}-\frac{E_{22}}{B^2r^2}=
{4\pi} (\mu+3P+\Pi_2)
-\frac{A^\prime}{B^2A}\left(\frac{2D^{\prime}}{D}
+\frac{B^{\prime}}{B}+\frac{1}{r}\right)\nonumber \\
&+&\frac{A_\theta}{AB^2r^2} \left(\frac{B_\theta}{B}-\frac{2D_\theta}{D}\right)-\frac{1}{B^2r^2}\frac{A_{\theta \theta}}{A},
\label{19}
\end{eqnarray}

\begin{eqnarray}
\mathcal{E}_3 =-\frac{E_{33}}{D^2}+\frac{E_{22}}{B^2r^2}=4\pi \Pi_3
-\frac{A^\prime}{B^2A}\left(\frac{D^{\prime}}{D}
-\frac{B^{\prime}}{B}-\frac{1}{r}\right)
-\frac{A_\theta}{AB^2r^2} \left(\frac{D_\theta}{D}+\frac{B_\theta}{B}\right)+
\frac{1}{B^2r^2}\frac{A_{\theta \theta}}{A}.
\label{20}
\end{eqnarray}
\end{widetext}

\section{Vanishing complexity factor conditions}
\label{C}
\begin{widetext}
\begin{eqnarray}
Y_{TF_1}&=&\frac{1}{B^2 r}\left[\frac{A^{\prime}_\theta}{A}-
\frac{A^{\prime} B_\theta}{AB}-\frac{A_\theta}{A} \left(\frac{B^{\prime}}{B}+\frac{1}{r}\right)\right]=0,
\label{YTF1}
\end{eqnarray}

\begin{eqnarray}
Y_{TF_2} &=&\frac{A^{\prime\prime}}{B^2A}
-\frac{A^\prime}{B^2A}\left(\frac{D^{\prime}}{D}
+\frac{B^{\prime}}{B}\right)+\frac{A_\theta}{AB^2r^2} \left(\frac{B_\theta}{B}-\frac{D_\theta}{D}\right)=0,
\label{YTF2}
\end{eqnarray}

\begin{eqnarray}
Y_{TF_3} &=&
-\frac{A^\prime}{B^2A}\left(\frac{D^{\prime}}{D}
-\frac{B^{\prime}}{B}-\frac{1}{r}\right)
-\frac{A_\theta}{AB^2r^2} \left(\frac{D_\theta}{D}+\frac{B_\theta}{B}\right)+
\frac{1}{B^2r^2}\frac{A_{\theta \theta}}{A}\nonumber \\&=&0.
\label{YTF3}
\end{eqnarray}
\end{widetext}
\section{Einstein equations for the dynamical spherically symmetric case}
 \label{D}
 Einstein's field equations
\begin{equation}
G_{\alpha\beta}=8\pi T_{\alpha\beta},
\label{2}
\end{equation}
 for the interior spacetime (\ref{1}) read

\begin{widetext}
\begin{eqnarray}
8\pi T_{00}=8\pi  \mu A^2
=\left(2\frac{\dot{B}}{B}+\frac{\dot{R}}{R}\right)\frac{\dot{R}}{R}
-\left(\frac{A}{B}\right)^2\left[2\frac{R^{\prime\prime}}{R}+\left(\frac{R^{\prime}}{R}\right)^2
-2\frac{B^{\prime}}{B}\frac{R^{\prime}}{R}-\left(\frac{B}{R}\right)^2\right],
\label{12} 
\end{eqnarray}
\end{widetext}
\begin{equation}
8\pi T_{01}=-8\pi qAB
=-2\left(\frac{{\dot R}^{\prime}}{R}
-\frac{\dot B}{B}\frac{R^{\prime}}{R}-\frac{\dot
R}{R}\frac{A^{\prime}}{A}\right),
\label{13} 
\end{equation}
\begin{widetext}
\begin{eqnarray}
8\pi T_{11}=8\pi P_r B^2 
=-\left(\frac{B}{A}\right)^2\left[2\frac{\ddot{R}}{R}-\left(2\frac{\dot A}{A}-\frac{\dot{R}}{R}\right)
\frac{\dot R}{R}\right]
+\left(2\frac{A^{\prime}}{A}+\frac{R^{\prime}}{R}\right)\frac{R^{\prime}}{R}-\left(\frac{B}{R}\right)^2,
\label{14} 
\end{eqnarray}

\begin{eqnarray}
8\pi T_{22}&=&\frac{8\pi}{\sin^2\theta}T_{33}=8\pi P_{\perp}R^2
=-\left(\frac{R}{A}\right)^2\left[\frac{\ddot{B}}{B}+\frac{\ddot{R}}{R}
-\frac{\dot{A}}{A}\left(\frac{\dot{B}}{B}+\frac{\dot{R}}{R}\right)
+\frac{\dot{B}}{B}\frac{\dot{R}}{R}\right]\nonumber \\
&+&\left(\frac{R}{B}\right)^2\left[\frac{A^{\prime\prime}}{A}
+\frac{R^{\prime\prime}}{R}-\frac{A^{\prime}}{A}\frac{B^{\prime}}{B}
+\left(\frac{A^{\prime}}{A}-\frac{B^{\prime}}{B}\right)\frac{R^{\prime}}{R}\right].\label{15}
\end{eqnarray}
\end{widetext}
The component (\ref{13}) can be rewritten with (\ref{5c1}) and
(\ref{5b1}) as
\begin{equation}
4\pi qB=\frac{1}{3}(\Theta-\sigma)^{\prime}
-\sigma\frac{R^{\prime}}{R}.\label{17a}
\end{equation}
\section{Dynamical equations}
\label{E}
The non trivial components of the Bianchi identities, $T^{\alpha\beta}_{;\beta}=0$, from (\ref{2}) yield

\begin{eqnarray}
T^{\alpha\beta}_{;\beta}V_{\alpha}&=&-\frac{1}{A}\left[\dot { \mu}+
\left( \mu+ P_r\right)\frac{\dot B}{B}
+2\left( \mu+P_{\perp}\right)\frac{\dot R}{R}\right] 
\nonumber \\&-&\frac{1}{B}\left[ q^{\prime}+2 q\frac{(AR)^{\prime}}{AR}\right]=0, \label{j4}
\end{eqnarray}
\begin{eqnarray}
T^{\alpha\beta}_{;\beta}\chi_{\alpha}=\frac{1}{A}\left[\dot { q}
+2 q\left(\frac{\dot B}{B}+\frac{\dot R}{R}\right)\right] \nonumber \\
+\frac{1}{B}\left[ P_r^{\prime}
+\left(\mu+ P_r \right)\frac{A^{\prime}}{A}
+2( P_r-P_{\perp})\frac{R^{\prime}}{R}\right]=0, \label{j5}
\end{eqnarray}

or, by using (\ref{5c}), (\ref{5c1}), (\ref{23a}) and (\ref{20x}), they become, respectively,
\begin{widetext}
\begin{eqnarray}
D_T \mu+\frac{1}{3}\left(3 \mu+ P_r+2P_{\perp} \right)\Theta 
+\frac{2}{3}( P_r-P_{\perp})\sigma+ED_R q
+2 q\left(a+\frac{E}{R}\right)=0, \label{j6}\\
D_T q+\frac{2}{3} q(2\Theta+\sigma)
+ED_R  P_r 
+\left( \mu+ P_r \right)a+2(P_r-P_{\perp})\frac{E}{R}=0.
\label{j7}
\end{eqnarray}
\end{widetext}
This last equation may be further tranformed as follows, the acceleration $D_TU$ of an infalling particle can
be obtained by using (\ref{5c}), (\ref{14}), (\ref{17masa})  and (\ref{20x}),
producing
\begin{equation}
D_TU=-\frac{m}{R^2}-4\pi  P_r R
+Ea, \label{28}
\end{equation}
and then, substituting $a$ from (\ref{28}) into
(\ref{j7}), we obtain
\begin{widetext}
\begin{eqnarray}
\left( \mu+ P_r\right)D_TU 
&=&-\left(\mu+ P_r \right)
\left[\frac{m}{R^2}
+4\pi  P_r R\right] 
-E^2\left[D_R  P_r
+2(P_r-P_{\perp})\frac{1}{R}\right] \nonumber \\
&-&E\left[D_T q+2 q\left(2\frac{U}{R}+\sigma\right)\right].
\label{3m}
\end{eqnarray}
\end{widetext}
\section{The complexity factors for the Bondi metric}
\label{F}
\begin{eqnarray}
{\cal E}_1=\frac{1}{r^2}\left(2 c_u \cot{\theta}+c_{\theta u}\right) +\mathcal{O}(r^{-n}), \quad n\geq 4,
\label{e1}
\end{eqnarray}
\begin{widetext}
\begin{eqnarray}
{\cal E}_2=\frac{1}{r}c_{uu}
-\frac{1}{2r^2}\left(c_{\theta \theta u} -4 M c_{u u}+2c_{u}+c_{\theta u}\cot{\theta}-\frac{4 c_{u}}{\sin^2{\theta}}\right)\nonumber \\
+\frac{1}{r^3}\left[c c_u +2 c_\theta c_{\theta u}+3M 
+\frac{\cot{\theta}}{2}\left(3 c_u c_\theta + 5 c c_{\theta u}\right) -M_u c + \frac{1}{2}M_{\theta \theta}+N_{\theta u} 
 +P_{uu}\right. \nonumber \\ 
 \left.- \cot{\theta}\left(M c_{\theta u}+\frac{1}{2}M_\theta +N_u -N c_{uu}\right)
 -M c_u\left(1-\frac{4}{\sin^2{\theta}}\right)+c_u\left(c c_u +\frac{1}{2}c_{\theta \theta}\right)
\right. \nonumber \\ 
\left. +c_{uu}\left(4M^2+N_{\theta}\right)-c_{\theta \theta u}\left(M-\frac{3}{2}c\right)\right] \nonumber \\+\mathcal{O}(r^{-n}), \nonumber \\ 
\label{e2a}
\end{eqnarray}
\end{widetext}
\begin{widetext}
\begin{eqnarray}
{\cal E}_3=\frac{2}{r}c_{uu}
-\frac{1}{r^2}\left(c_{\theta \theta u} -4 M c_{u u}+2c_{u}+c_{\theta u}\cot{\theta}-\frac{4 c_{u}}{\sin^2{\theta}}\right)
\nonumber \\
+\frac{1}{r^3}\left[-4 c c_u +4 c_\theta c_{\theta u} 
 +\cot{\theta}\left(3c_u c_\theta + 5c c_{\theta u}\right) -2M_u c + M_{\theta \theta}+2N_{\theta u} +2P_{uu}\right.\nonumber \\
\left. -\cot{\theta}\left(2M c_{\theta u}+M_\theta +2N_u -2N c_{uu}\right)
 -2M c_u \left(1-\frac{4}{\sin^2{\theta}}\right)+c_u\left(2c c_u +c_{\theta \theta}\right)\right. \nonumber\\
\left. +2c_{uu}\left(4M^2+N_{\theta}\right)-c_{\theta \theta u}\left(2M-3c\right)\right] \nonumber  \\ +\mathcal{O}(r^{-n}).  \nonumber \\
\label{e3}
\end{eqnarray}
\end{widetext}

\section{The magnetic part of the Weyl tensor}
\label{G}
\begin{eqnarray}
H_1=-\frac{1}{r^2} \left(2 c_u \cot{\theta}+c_{\theta u}\right) +\mathcal{O}(r^{-n}), \quad n\geq 4,
\label{h1}
\end{eqnarray}
\\
\begin{widetext}
\begin{eqnarray}
H_2=-\frac{1}{r}c_{uu}-\frac{1}{r^2}\left[-c_u\left(1-\frac{2}{\sin^2{\theta}}\right)
 -\frac{\cot{\theta}}{2}c_{\theta u}+ 2 c_{uu}(M-c)-\frac{1}{2}c_{\theta \theta u}\right]\nonumber\\
-\frac{1}{r^3}\left\{-Mc_u\left(1-\frac{4}{\sin^2{\theta}}\right) - \frac{4cc_u}{\sin^2{\theta}}
+\cot{\theta}\left[\frac{3}{2}c_u c_\theta - N_u-\frac{1}{2}M_{\theta}+N c_{uu}+\left(\frac{7}{2}c-M\right)c_{\theta u}\right]+\left(\frac{5}{2}c-M\right)c_{\theta \theta u}\right. \nonumber \\
\left.+\frac{1}{2}c_{\theta \theta} c_{u}+2c_{\theta}c_{\theta u}+c c_{u}^2+\frac{1}{2}M_{\theta \theta}-cM_{u}
+N_{\theta u}+P_{uu}+c_{uu}\left(4c^2+4M^2-4Mc+N_{\theta}\right)\right\}+\mathcal{O}(r^{-n}),\quad  n\geq 4\nonumber\\
\label{h2}
\end{eqnarray}
\end{widetext}

where $P=C-\frac{c^3}{6}$.
\section{The vorticity}
\label{H}
\begin{widetext}
\begin{eqnarray}
\omega^{\phi}  =  -\frac{e^{-2\beta}}{2 r^2 \sin{\theta}} \left[ 2
\beta_\theta e^{2\beta} - \frac{2 e^{2\beta} A_\theta}{A}
- \left(U r^2 e^{2\gamma}\right)_r  
  +  \frac{2 U r^2 e^{2\gamma}}{A} A_r +
\frac{e^{2\beta}\left(U r^2 e^{2\gamma}\right)_u}{A^2} - \frac{Ur^2
e^{2\gamma}}{A^2} 2 \beta_u e^{2\beta} \frac{}{} \right], \label{om3}
\end{eqnarray}
\end{widetext}
and for the absolute value of $\omega^\alpha$ we get
\begin{widetext}
\begin{eqnarray}
\Omega  \equiv  \left(- \omega_\alpha \omega^\alpha\right)^{1/2} =
  \frac{e^{-2\beta -\gamma}}{2 r}
\left[2 \beta_\theta e^{2\beta} - 2 e^{2\beta} \frac{A_\theta}{A}
  -  \left(U r^2 e^{2\gamma}\right)_r +  2 U r^2 e^{2\gamma} \frac{A_r}{A}+  \frac{e^{2\beta}}{A^2} \left(U r^2 e^{2\gamma}\right)_u
  - 2 \beta_u \frac{e^{2\beta}}{A^2} U r^2 e^{2\gamma}
\right] \label{OM}
\end{eqnarray}
\end{widetext}
Feeding back (\ref{ga}--\ref{be}) into (\ref{OM}) and
keeping only the two leading terms, we obtain
\begin{widetext}
\begin{eqnarray}
\Omega  = -\frac{1}{2r} ( c_{u \theta}+2 c_u \cot \theta)  +\frac 1{r^2} \left[ M_{\theta}-M (c_{u \theta}+2 c_u \cot 
\theta)-c c_{u
\theta}+6 c c_u \cot \theta+2 c_u c_{\theta} \right].
\label{Om2}
\end{eqnarray}
\end{widetext}


\end{document}